\documentclass[a4paper,11pt]{article}
\pdfoutput=1 
\usepackage{jheppub} 
\usepackage{bbm,bm,graphicx,slashed,mathtools,bbold}
\usepackage[T1]{fontenc}

\DeclareMathOperator\diag{diag}
\DeclareMathOperator\ee{e}
\DeclareMathOperator\Tr{Tr}

\renewcommand\bar[1]{\overline{#1}}
\renewcommand\epsilon{\varepsilon}
\renewcommand\phi{\varphi}

\newcommand\1{\mathbb{1}}
\newcommand\HH{\bm{H}}
\newcommand\Mm{\hat{M}}
\newcommand\Mu{\bm{\mu}}
\newcommand\Muu{\hat{\bm{\mu}}}
\newcommand\Nf{N_f}
\newcommand\Nfhalf{N_f/2}
\newcommand\Pis{\hat\Omega}
\newcommand\SB{{high-isospin-density Silver Blaze phenomenon}}
\newcommand\SU{\text{SU}}
\newcommand\Sp{\text{Sp}}
\newcommand\U{\text{U}}
\newcommand\YM{\text{YM}}
\newcommand\dMu{\check{\Mu}}
\newcommand\del{\partial}
\newcommand\dmu{\check\mu}
\newcommand\mL{\mathcal{L}}
\newcommand\mm{\hat{m}}
\newcommand\mui{\mu_I}
\newcommand\muu{\hat{\mu}}
\newcommand\pis{\Omega}
\newcommand\piss{\hat{\omega}}
\newcommand\pisss{\omega}
\newcommand\pq{phq}
\newcommand\qcd{\text{QCD}}
\newcommand\rmt{\text{RMT}}

\preprint{RIKEN-QHP-157}

\title{Stressed Cooper pairing in QCD at high isospin density: 
effective Lagrangian and random matrix theory}

\author[a]{Takuya Kanazawa}
\author[b]{and Tilo Wettig}

\affiliation[a]{iTHES Research Group and Quantum Hadron Physics
  Laboratory, RIKEN, Wako, Saitama 351-0198, Japan}
\affiliation[b]{Department of Physics, University of Regensburg, 93040
  Regensburg, Germany}

\emailAdd{takuya.kanazawa@riken.jp}
\emailAdd{tilo.wettig@ur.de}

\abstract{We generalize QCD at asymptotically large isospin chemical
  potential to an arbitrary even number of flavors.  We also allow for
  small quark chemical potentials, which stress the coincident Fermi
  surfaces of the paired quarks and lead to a sign problem in Monte
  Carlo simulations.  We derive the corresponding low-energy effective
  theory in both $p$- and $\epsilon$-expansion and quantify the
  severity of the sign problem.  We construct the random matrix theory
  describing our physical situation and show that it can be mapped to
  a known random matrix theory at low baryon density so that new insights can
  be gained without additional calculations.  In particular, we
  explain the Silver Blaze phenomenon at high isospin density.  We
  also introduce stressed singular values of the Dirac operator and
  relate them to the pionic condensate.  Finally we comment on
  extensions of our work to two-color QCD.}

\begin{document} 
\maketitle
\flushbottom

\section{Introduction}
\label{sec:intro}

Understanding the nonperturbative physics of Quantum Chromodynamics
(QCD) is one of the central challenges in theoretical physics. QCD in
the vacuum is strongly coupled, giving rise to a variety of emergent
phenomena such as chiral symmetry breaking, quark confinement,
formation of nuclei, and mass gap generation of gluons. Since the
seminal work by Banks and Casher \cite{Banks:1979yr} it is known that
chiral symmetry breaking is associated with the condensation of
near-zero eigenvalues of the Dirac operator. The correlations of Dirac
eigenvalues on the scale $\sim 1/V_4\Sigma$, also known as the
microscopic domain, strictly obey the predictions of chiral random
matrix theory (ChRMT), which corresponds to the leading order of the
$\epsilon$-expansion of chiral perturbation theory (ChPT)
\cite{Shuryak:1992pi,Verbaarschot:1993pm} (see
\cite{Verbaarschot:1997bf,Verbaarschot:2000dy} for reviews). Here,
$V_4$ and $\Sigma$ stand for the volume of Euclidean space-time and
the chiral condensate in the chiral limit, respectively. The
equivalence between a rather simple Gaussian matrix model with no
space-time structure and QCD in a certain limit is truly surprising,
but it has been confirmed explicitly again and again through lattice
QCD simulations. Not only theoretically intriguing, the equivalence
also provides us with a means of extracting low-energy constants in
ChPT from lattice QCD data, where Dirac eigenvalues are easily
computable.

The dynamics of QCD at nonzero temperature $T$ and/or chemical
potential $\mu$ is relevant for the physics of the early Universe,
relativistic heavy-ion collisions, and compact stars
\cite{Kogut:2004su,Yagi:2005yb,Fukushima:2010bq}. At high baryon density, the
physics is entirely different from that of the vacuum: the celebrated
BCS mechanism leads to the condensation of quark pairs, which breaks
gauge and chiral symmetries in three-color QCD, a phenomenon referred
to as color superconductivity
\cite{Rajagopal:2000wf,Alford:2007xm}. However, conventional Monte
Carlo simulations based on importance sampling are hindered by the
infamous sign problem, which originates from the complex phase of the
fermion determinant at nonzero $\mu$ \cite{deForcrand:2010ys}. While
several promising approaches to overcome this obstacle have been
proposed \cite{deForcrand:2010ys,Sexty:2013ica}, a feasible way to
simulate dense QCD is yet to be found. To gain insights into the
physics of dense quark matter, a number of QCD-like theories that have
a nonnegative path-integral measure even at nonzero chemical potential
have been investigated intensively by many authors, with numerical
methods as well as in effective models. Such special theories include
QCD with gauge group $\SU(2)$ (called two-color QCD)
\cite{Hands:1999md}, QCD with adjoint fermions \cite{Hands:2000ei},
$G_2$ gauge theory \cite{Maas:2012ts}, and QCD with isospin chemical
potential \cite{Alford:1998sd,Son:2000xc}.%
\footnote{In two-color QCD, the positivity of the measure is ensured
  for an even number of flavors and pairwise equal masses. In QCD with
  adjoint fermions, no such restriction is necessary.} Those theories
share many features, such as the existence of light bosons that
condense at nonzero chemical potential, and the interested reader is
referred to \cite{vonSmekal:2012vx,Kanazawa:PhD} for reviews.

In the absence of reliable numerical simulations, analytical
first-principle studies are highly valuable. The study of the Dirac
spectrum in QCD with nonzero quark chemical potential in the regime
$\mu^2_q \ll 1/\sqrt{V_4}$ was undertaken in
\cite{Osborn:2004rf,Akemann:2004dr,Osborn:2005ss,Osborn:2008jp,Osborn:2008ab}
on the basis of low-energy effective theories and ChRMT (see
\cite{Akemann:2004dr,Verbaarschot:2005rj} for reviews). It was found
that the sign problem is manifested in an extreme oscillation of the
spectral density of the Dirac operator, and that the latter is
actually responsible for the fact that observables in QCD (e.g., the
chiral condensate) at $T=0$ are independent of $\mu_q$ below roughly
one third of the nucleon mass, even though the fermion determinant
itself depends on $\mu_q$. This is informally called the Silver Blaze
phenomenon of QCD \cite{Cohen:2003kd,Cohen:2004qp}. The baryon-number
Dirac spectrum was also studied in \cite{Ipsen:2012ug}.

The microscopic Dirac spectrum in QCD and QCD-like theories at high
density was investigated in
\cite{Yamamoto:2009ey,Kanazawa:2009en,Kanazawa:2009ks,Akemann:2010tv,Kanazawa:PhD}. Through
the extension of ChRMT to dense QCD it was shown that the fluctuations
of the complex Dirac eigenvalues of order $1/\sqrt{V_4\Delta^2}$ (with
$\Delta$ the BCS gap of quarks) are universal, i.e., independent of
the microscopic details of the QCD interaction and solely determined
by global symmetries.  The whole analysis was extended to the singular
values of the Dirac operator \cite{Kanazawa:2011tt}. A
Banks-Casher-type relation in dense QCD-like theories was also
established, which connects the Dirac spectral density at the origin
and $\Delta^2$ \cite{Kanazawa:2012zr}.

In this paper we consider QCD with an even number $N_f$ of flavors at
asymptotically large isospin chemical potential $\mui\gg\Lambda_{\qcd}$
\cite{Son:2000xc,Son:2000by}. For two flavors and zero quark chemical
potential the partition function is given by%
\footnote{In this work we define $\mui$ as $-1/2$ times the
  conventional isospin chemical potential so that $\mui>0$ leads 
  to a finite density of $\bar u$ and $d$ quarks. See also footnote \ref{fn:mui}.}
\begin{align}
  \label{eq:Z_11}
  Z_{\qcd}^{(N_f=2)}(\mui) & = \Big\langle {\det} [D(-\mui+\mu_q)+m ] ~
  {\det}[D(\mui+\mu_q)+m] \Big\rangle_{\YM}\bigg|_{\mu_q=0}
  \\
  \label{eq:Z_1+1}
  & = \Big\langle \big| {\det}[D(\mui)+m] \big|^2 
  \Big\rangle_{\YM} 
  \,,
\end{align}
where the Dirac operator $D$ with the property $D(\mu)^\dagger =
-D(-\mu)$ is defined in section~\ref{sec:prelim} and the subscript
$\YM$ implies an average over the gauge fields. At low $T$, the ground
state is dominated by the Fermi sea of $\bar u$ and $d$ quarks plus
the condensate $\langle \bar u\gamma_5 d \rangle$ that originates from
the attractive interaction between quarks near the Fermi
surface.\footnote{The latter is also supported by a QCD inequality
  \cite{Son:2000xc}.} This leads to a BCS gap $\Delta$ for quarks. In
\cite{Kanazawa:PhD} a low-energy effective theory at energy scale $\ll
\Delta$ was constructed for the generalization of \eqref{eq:Z_1+1} to
$N_f$ flavors. Furthermore, the ChRMT describing the spectrum of
$D(\mu)$ was identified and solved analytically
\cite{Kanazawa:PhD}.

\begin{figure}[tb]
  \centering
  \includegraphics[width=10.5cm]{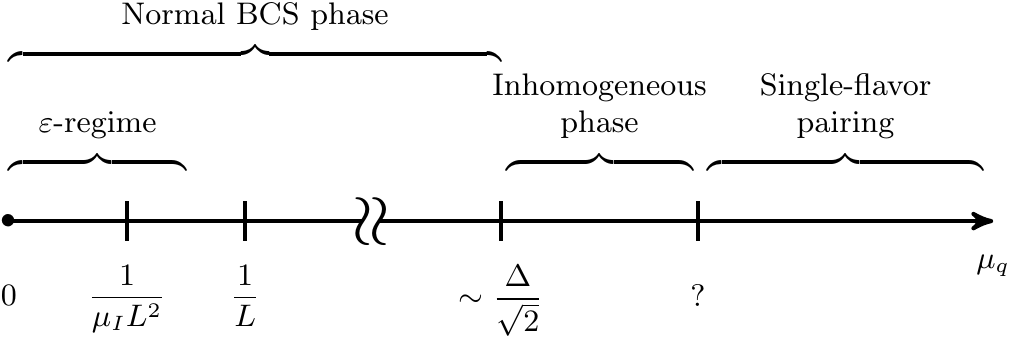}
  \caption{\label{fig:schematic} A schematic phase diagram of $N_f=2$
    QCD with large isospin chemical potential $\mui$ as a function of
    quark chemical potential $\mu_q$ at $T=0$. The $\epsilon$-regime
    will be defined in section~\ref{sec:LEET} assuming that the system
    is placed in a four-dimensional Euclidean box of linear extent
    $L$.  }
\end{figure}

Then a natural question to ask is what happens if the condition
$\mu_q=0$ is loosened.  This is a long-standing subject, and a rough
physical picture is known at least for asymptotically large $\mui$
where the weak-coupling BCS mechanism is at work (see
figure~\ref{fig:schematic}). Namely, for small $\mu_q \ne 0$, the
pairing between $\bar u$ and $d$ quarks is stressed by the mismatch of
Fermi levels, but the ground state at $T=0$ is unchanged as long as
$\mu_q$ is too small to compensate for the energy cost of breaking the
Cooper pairs. In this region, no quark number is generated and 
the BCS gap is independent of $\mu_q$ \cite{Bedaque:1999nu}. 
(This property will be referred to as the {\SB}
in the rest of this paper, to distinguish it from the original one at
low baryon density.) When $\mu_q$ reaches a threshold $\mu_q^c\approx
\Delta/\sqrt{2}$ (called the Chandrasekhar-Clogston limit
\cite{Clogston:1962zz,Chandrasekhar1962}) the standard BCS pairing is
no longer energetically preferable and a phase transition
occurs to an inhomogeneous phase (e.g., a
Fulde-Ferrell-Larkin-Ovchinnikov phase, where the pair carries a net
nonzero momentum) \cite{Alford:2000ze}. As $\mu_q$ grows further, the
system is expected to undergo yet another phase transition to a state
with a single-flavor pairing ($\bar u \,\bar u$ and $dd$)
\cite{Schafer:2000tw,Schmitt:2004et}. On the other hand, for low and
intermediate $\mui$ the physics is less transparent because the system
is strongly coupled; see, e.g.,
\cite{Splittorff:2000mm,Kogut:2001id,Kiriyama:2001ud,Klein:2003fy,
  Nishida:2003fb,Barducci:2004tt,Loewe:2005yn,He:2005nk,Lawley:2005ru,
  Ebert:2005wr,He:2006tn,Mao:2006zr,Mukherjee:2006hq,
  Fukushima:2007bj,Andersen:2007qv,Nickel:2008ng,Andersen:2010vu,
  Sasaki:2010jz,Mu:2010zz,Kamikado:2012bt,Abuki:2013vwa,Ueda:2013sia,
  Stiele:2013pma,Xia:2013caa,Kang:2013bea,He:2013gga,Nishihara:2014nva}
for studies on QCD-like theories and
\cite{Casalbuoni:2003wh,Anglani:2013gfu,Buballa:2014tba} for reviews 
on possible inhomogeneous phases in the phase diagram. 
We also note that in recent years similar physics has been discussed
in the context of imbalanced ultracold atomic Fermi gases
\cite{Sheehy:2006qc,Chevy:2010zz}.

The {\SB} for $0<\mu_q<\mu_q^c$ is puzzling at first sight, since
observables are independent of $\mu_q$ while the fermion determinant
${\det} [D(-\mui+\mu_q)+m ] ~{\det}[D(\mui+\mu_q)+m]$ in the
path-integral measure depends on $\mu_q$. In this paper we elucidate
the mechanism behind this phenomenon by constructing the low-energy
effective theory and the corresponding ChRMT for QCD at large isospin
and small quark chemical potential, and by looking into the spectral
properties of the Dirac operator. As an idealization we will neglect 
beta decay and the charge neutrality condition, which would 
strongly suppress the formation of a pionic condensate 
\cite{Ebert:2005wr,Andersen:2007qv,Abuki:2008tx}. 

This paper is organized as follows.  In section~\ref{sec:prelim} we
summarize basic properties of QCD with isospin and quark chemical
potential and fix the notation.  In section~\ref{sec:LEET} we
construct the corresponding low-energy effective theory in both $p$-
and $\epsilon$-expansion and study the severity of the sign problem.
In section~\ref{sec:rmt} we construct the ChRMT corresponding to the
leading order of the $\epsilon$-expansion.  By mapping it to a known
ChRMT that is applicable at low baryon density we can gain a number of
insights at high isospin density.  We define stressed singular values of the
Dirac operator and relate them to the pionic condensate, and also
study the baryon-number Dirac spectrum.  In section~\ref{sec:2color}
we briefly comment on two-color QCD and mention which parts of the
arguments for QCD with $N_c\geq 3$ have to be modified for $N_c=2$.
We conclude in section~\ref{sec:conclusion}.  In
Appendix~\ref{app:red} we clarify a potential ambiguity in the
effective theory.

\section{QCD with large isospin chemical potential}
\label{sec:prelim}

\begin{figure}[t] 
  \centerline{\includegraphics[width=.7\textwidth]{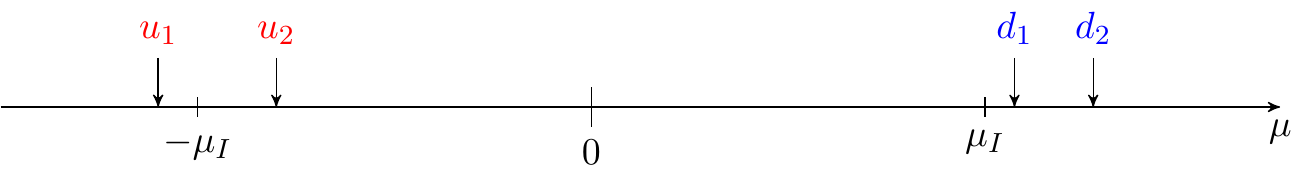}}
  \caption{\label{fg:stressed} A typical situation considered in this
    paper is shown for $N_f=4$.  The chemical potentials for $u$
    quarks ($d$ quarks) are assumed to be slightly perturbed from
    $-\mui$ ($+\mui$).}
\end{figure}

Assuming even $\Nf$, we consider QCD with $\Nfhalf$ pairs of $u$ and
$d$ quarks.  We will refer to $u$ quarks as $u_f$ and to $d$ quarks as
$d_f$ with $f=1,\dots,N_f/2$.  We introduce chemical potentials of the
form $\mu_{u,f}=-\mui+\dmu_{u,f}$ for the $u$ quarks and
$\mu_{d,f}=\mui+\dmu_{d,f}$ for the $d$ quarks, respectively, where we
assume $|\dmu_{i,f}|\ll\mui$ for $i=u,d$ and all $f$.  In other words,
we consider QCD at large isospin chemical potential $\mui$ but allow
for small quark chemical potentials on top of $\mui$.  For convenience
of notation we define
\begin{subequations}
  \begin{alignat}{2} 
    \bm\mu_u & =-\mui\1_{\Nfhalf}+\dMu_u & \quad \text{with}\quad 
    \dMu_u&=\diag(\dmu_{u,1},\ldots,\dmu_{u,\Nfhalf})\,,
    \\
    \bm\mu_d & =\mui\1_{\Nfhalf}+\dMu_d & \quad \text{with} \quad 
    \dMu_d&=\diag(\dmu_{d,1},\ldots,\dmu_{d,\Nfhalf})\,.  
  \end{alignat}
\end{subequations}
An example of a chemical potential distribution for $N_f=4$ is shown
in figure~\ref{fg:stressed}.

The partition function of the microscopic theory is given by
\begin{align}
  & Z^{(\Nf)}_{\qcd}(\mui; \{\dmu\},\{m\}) \notag\\
  \label{eq:Z_gen}
  &\qquad = \Big\langle \prod_{f=1}^{\Nfhalf}
  \underbrace{\det\big(D(-\mui+\dmu_{u,f})+m_{u,f}\big)}_\text{$u$ quarks}\,
  \underbrace{\det\big(D(\mui+\dmu_{d,f})+m_{d,f}\big)}_\text{$d$ quarks}
  \Big\rangle_{\YM}\,,
\end{align}
where\footnote{\label{fn:mui}In the literature one sometimes finds the definition
  $D(\mu)=\gamma_\nu D_\nu + \mu\gamma_4$, which interchanges the
  meaning of positive and negative $\mu$.  With our current
  definition, a positive $\mu$ favors quarks over anti-quarks.  For
  convenience of notation we choose $\mui>0$, i.e., assigning $-\mui$
  ($+\mui$) to $u$ ($d$) quarks favors $\bar ud$ over $\bar du$.}
$D(\mu)\equiv\gamma_\nu D_\nu - \mu\gamma_4$ is the Euclidean Dirac
operator in the fundamental representation of $\SU(N_c)$ for $N_c\geq
3$,\footnote{The special case $N_c=2$ will be discussed briefly in
  section~\ref{sec:2color}.} which is an analytic continuation of the
Minkowski Dirac operator $D_M(\mu)\equiv i\gamma^\nu D_\nu +\mu
\gamma^0$ with $x^0=-ix_4$. In this paper we always work in Euclidean
space-time unless stated otherwise.  We also assume sufficiently low
temperature $T\ll \Delta$ throughout. The special case
\eqref{eq:Z_11} is recovered by setting $\Nf=2$,
$\dmu_{u,1}=\dmu_{d,1}=\mu_q$, and $m_{u,1}=m_{d,1}=m$ in
\eqref{eq:Z_gen}.

Note that shifting $\dMu_u \to \dMu_u -\delta\dmu\1_{\Nfhalf}$ and
$\dMu_d \to \dMu_d + \delta\dmu\1_{\Nfhalf}$ simply corresponds to a
shift $\mui \to \mui+\delta\dmu$, as is evident from \eqref{eq:Z_gen}.
We are not interested in such a trivial shift and therefore impose the
condition
\begin{align}
  \label{eq:GF}
  \Tr[\dMu_u-\dMu_d]=0\,,
\end{align}
which implies $\mui=\Tr[\bm\mu_d-\bm\mu_u]/\Nf$.  Not imposing this
condition leads to an ambiguity in the effective theory that is
discussed in appendix~\ref{app:red}, which should best be read after
section~\ref{sec:nonzerostress}.

\section{Low-energy effective theory}
\label{sec:LEET}

\subsection[\texorpdfstring{$p$-expansion}{p-expansion}]{\boldmath
  $p$-expansion}

The purpose of this subsection is to derive a low-energy effective
theory of Nambu-Goldstone (NG) bosons for the theory defined in
section~\ref{sec:prelim}.  We are interested in a regime where the
coincident Fermi surfaces of $\bar{u}$ and $d$ quarks are slightly disrupted
by nonzero $|\dmu_{i,f}|\ll \Delta$. Before looking into this case we
first consider the limit $\dMu_u=\dMu_d=0$
\cite{Son:2000xc,Son:2000by} as a starting point.

\subsubsection{Effective theory for zero stress}
\label{sec:zerostress}

As noted in \cite{Son:2000xc,Son:2000by,Kanazawa:PhD}, the symmetry
breaking for \eqref{eq:Z_gen} at $\mui\gg\Lambda_{\qcd}$ (and
$\dMu_u=\dMu_d=0$) is driven by the condensate\footnote{This
  pseudoscalar channel is favored over the scalar channel by positive
  quark masses and the instanton-induced interactions
  \cite{Son:2000xc}.} $\langle \bar u_f \gamma_5 d_f \rangle$
$(f=1,\dots,\Nfhalf)$, resulting in the breaking pattern
\cite[Sec.~4.2]{Kanazawa:PhD}
\begin{align}
  &\U(\Nfhalf)_{u_R} \times \U(\Nfhalf)_{u_L} \times 
  \U(\Nfhalf)_{d_R} \times \U(\Nfhalf)_{d_L} 
  \notag\\
  &\longrightarrow  ~~
  \U(\Nfhalf)_{u_R+d_L} \times \U(\Nfhalf)_{u_L+d_R}\,,
  \label{eq:SSB}
\end{align}
where the suppression of the axial anomaly by medium effects is taken
into account. This pattern of spontaneous symmetry breaking is
consistent with QCD inequalities \cite{Kanazawa:PhD}.%
\footnote{For general mass terms and nonzero $\dMu_{u,d}$, the
  path-integral measure of QCD is not necessarily positive definite
  and QCD inequalities do not apply.  In that case more exotic pairing
  patterns are possible, but only if the masses or $\dMu_{u,d}$ are
  large enough.  Here we assume them to be small perturbations so that
  \eqref{eq:SSB} remains valid.}

The breaking pattern \eqref{eq:SSB} gives rise to $N_f^2/2$ NG bosons
which we parameterize by $U$ and $V$ and which reside in the coset
spaces
\begin{align}
  U\in \frac{\U(\Nfhalf)_{u_R} \times \U(\Nfhalf)_{d_L}}
  {\U(\Nfhalf)_{u_R+d_L}}\cong\U(\Nfhalf)\,, \quad
  V\in \frac{\U(\Nfhalf)_{u_L} \times \U(\Nfhalf)_{d_R}}
  {\U(\Nfhalf)_{u_L+d_R}}\cong\U(\Nfhalf)\,.
\end{align} 
We can employ the spurion method to determine the form of the
effective Lagrangian $\mathcal L_\text{eff}$.  Under flavor
transformations of quarks, the NG bosons and the quark masses
transform as
\begin{align}
  \label{eq:MUV}
  M_u \to g_{u_L}M_u \,g_{u_R}^\dagger\,,\quad 
  M_d \to g_{d_L}M_d \,g_{d_R}^\dagger\,,\quad 
  U \to g_{d_L} U \,g_{u_R}^\dagger\,,\quad
  V \to g_{u_L} V \,g_{d_R}^\dagger\,,
\end{align}
where $M_u$ and $M_d$ are the mass matrices for the $u$ and $d$
quarks\footnote{Equation \eqref{eq:Z_gen} has been expressed in a
  basis in which the mass matrices are diagonal, i.e.,
  $M_u\to\diag(m_{u,1},\ldots,m_{u,\Nfhalf})$ and
  $M_d\to\diag(m_{d,1},\ldots,m_{d,\Nfhalf})$.} and $g_i\in
\U(\Nfhalf)_i$ for $i\in \{u_{R},u_L,d_R,d_L\}$.

For later use, let us also insert a source term 
\begin{align}
  \label{eq:sce}
  \bar u_f \big[ (\pis_1)_{fg} P_L + (\pis_2)_{fg} P_R \big] d_g 
  +\text{h.c.}
\end{align}
into the Lagrangian, where $\pis_1$ and $\pis_2$ are
$(\Nfhalf)\times(\Nfhalf)$ matrices and $P_{R/L} = (\1\pm\gamma_5)/2$
are the usual chiral projectors.  This allows us to extract the pionic
condensate $\langle \bar u\gamma_5 d \rangle+\text{c.c.}$ by taking
the derivative of $\log Z_{\qcd}^{(\Nf)}$ w.r.t.\ $\pis_{1,2}$. The
role of this source term is similar to that of the diquark source in
two-color QCD and adjoint QCD. This term enables us to derive a
Banks-Casher-type relation for the pionic condensate
\cite{Kanazawa:2011tt}. To leave \eqref{eq:sce} invariant under flavor
transformations, $\pis_1$ and $\pis_2$ should transform as
\begin{align}
  \label{eq:pis}
  \pis_1 \to g_{u_R} \pis_1 g_{d_L}^\dagger
  \quad\text{and}\quad
  \pis_2 \to g_{u_L} \pis_2 g_{d_R}^\dagger \,.
\end{align}
Next we consider the parity transformation $P$. Recalling $U\sim
d_L\bar{u}_R$ and $V\sim u_L\bar{d}_R$, we have
\begin{align}
  P: \quad U \to V^\dagger, \quad V\to U^\dagger, \quad 
  M_{u,d} \to M^\dagger_{u,d}\,,  \quad 
  \pis_1 \to \pis_2\,, \quad  \pis_2\to \pis_1\,.
\end{align}
Assuming the ``$p$-regime'' counting of this theory to be%
\footnote{We explain the reason for this counting
  after~\eqref{eq:L_eff}. Note that this counting only applies at high
  isospin density and must not be confused with the usual $p$-expansion in the
  vacuum, where $\del_\nu \sim O(p)$ and $M_{u,d} \sim O(p^2)$. }
\begin{align}
  \del_\nu\sim M_{u,d} \sim O(p) \quad \text{and} \quad \pis_{1,2} \sim O(p^2)\,, 
  \label{eq:pcounting}
\end{align}
the leading $O(p^2)$ effective Lagrangian invariant under
\eqref{eq:MUV}, \eqref{eq:pis}, and parity turns out to be
\begin{align}
  \label{eq:L_eff}
  \mathcal{L}_\text{eff}(U,V) = &\ 
  \frac{F^2}{4}\left\{
    \Tr[ \del_4 U^\dagger\del_4 U + v^2 \del_i U^\dagger\del_i U ] + 
    \Tr[ \del_4 V^\dagger\del_4 V + v^2 \del_i V^\dagger\del_i V ] 
  \right\}
  \notag
  \\
  & - f^2 \left\{(\Tr[U^\dagger \del_4 U])^2 
    +\tilde v^2 (\Tr[U^\dagger \del_i U])^2
    + (\Tr[V^\dagger \del_4 V])^2 
    +\tilde v^2 (\Tr[V^\dagger \del_i V])^2 \right\}  
  \notag 
  \\
  & + \frac{3N_c}{4\pi^2}\Delta^2 
  \big\{\!\Tr[M_uU^\dagger M_d V^\dagger] 
  + \text{c.c.} \big\} 
  - \Phi \, \big\{\!\Tr[ \pis_1 U + \pis_2 V^\dagger ] 
  + \text{c.c.} \big\} 
  \,.  
\end{align}
To better understand this result we add a few comments:
\begin{enumerate}
\item The non-existence of $O(M_{u,d})$ terms in
  $\mathcal{L}_\text{eff}(U,V)$ points to the fact that the chiral
  condensate vanishes in this theory owing to the huge energy gap of
  anti-quarks due to the Fermi sea.  It then follows from the first
  term on the third line of \eqref{eq:L_eff} that the masses of the NG
  modes are $m_\text{NG}^2=O(M_{u,d}^2)$.  In contrast, the sources
  $\pis_{1,2}$ appear linearly, as they couple to the condensate in
  this theory. This implies for the masses of the NG modes that
  $m_{\rm NG}^2 =O(\pis_{1,2})$. To perform a consistent low-energy
  expansion based on a propagator $1/(p^2+m_{\rm NG}^2)$ it is natural
  to count $M_{u,d}$ as $O(p)$ and $\pis_{1,2}$ as $O(p^2)$, which
  explains \eqref{eq:pcounting}.
\item Cross terms, i.e., $\Tr[U^\dagger \del_4 U]\Tr[V^\dagger \del_4
  V]$ and $\Tr[U^\dagger \del_i U]\Tr[V^\dagger \del_i V]$, are
  suppressed at high density \cite{Son:1999cm} and have been dropped
  here. Terms with a single derivative, i.e., $\Tr[U^\dagger \del_4
  U]$ and $\Tr[V^\dagger \del_4 V]$, are also allowed by symmetries,
  but these are total derivatives that do not contribute to the
  action. We note in passing that the second line in \eqref{eq:L_eff}
  only affects the $\U(1)$ part of $U$ and $V$ since $\Tr[{\tilde
    U}^\dagger \del\tilde U]=0$ for any $\tilde U\in\SU(\Nfhalf)$.
\item The low-energy constants in \eqref{eq:L_eff} are defined in the
  limit $M_u=M_d=\pis_1=\pis_2=0$ and depend on $\mui$ and $\Nf$.  $\Phi$ is
  proportional to the magnitude of the pionic condensate $\langle\bar
  u\gamma_5 d\rangle+\text{c.c.}$ $\Delta$ is the BCS gap of quarks.
  $F$ and $f$ are the decay constants of the NG modes, and $v$ and
  $\tilde v$ are the corresponding velocities in the medium.  At
  asymptotically high density we have relations such as
  $\Lambda_{\qcd}\ll \Delta\ll \mui$, $v = 1/\sqrt{3}$, $F\sim \mui$,
  and $\Phi \sim \mui^2\Delta/g$ \cite{Alford:2007xm},\footnote{These
    relations were originally derived for quark chemical potential,
    but the same techniques can be used to show that they are also
    valid for isospin chemical potential.} but precise knowledge of
  these quantities is not needed in the rest of this paper.
\item The coefficient $3N_c\Delta^2/4\pi^2$ of the first term on the
  third line of \eqref{eq:L_eff} was determined in
  \cite{Kanazawa:PhD,Kanazawa:2012zr} through matching between
  high-density effective theory (HDET)
  \cite{Hong:1998tn,Hong:1999ru,Schafer:2001za} and chiral effective
  theory (see \cite{Son:1999cm,Son:2000tu,Schafer:2001za} for the
  corresponding analysis in the color-flavor-locked phase).
  The positive overall sign of this term fixes the parity of the
  ground state: since the minimum of this term is attained at
  $U=-V\propto \1$ for $M_{u,d}$ real and positive, the ground state
  is odd under parity \cite{Son:2000xc}. If $\pis_1=-\pis_2$ (a source
  for the $0^-$ condensate) the last term of \eqref{eq:L_eff} is also
  minimized by $U=-V$. However, there will be a competition if
  $\pis_1=\pis_2$.
\item The so-called Bedaque-Sch\"{a}fer terms \cite{Bedaque:2001je}
  are not included in \eqref{eq:L_eff} as they are subleading in the
  present $p$-expansion.
\item Constant terms $\sim \Tr[M_i^\dagger M_i]\ (i=u,\,d)$ are not
  explicitly shown in \eqref{eq:L_eff} because they do not affect the
  dynamics of NG modes and because they are irrelevant for the
  analysis of microscopic Dirac eigenvalues \cite{Kanazawa:2012zr}.
\end{enumerate}

\subsubsection{Effective theory for nonzero stress}
\label{sec:nonzerostress}

We now incorporate the effects of $\dMu_{u,d}$ into \eqref{eq:L_eff},
assuming that these chemical potentials are much smaller than the gap
$\Delta$ and can thus be regarded as low-energy expansion parameters
in $\mathcal{L}_\text{eff}$.  For this purpose we again employ the
$p$-counting
\begin{align}
  \label{eq:p-rule}
  \del_\nu \sim M_{u,d} \sim \dMu_{u,d} \sim O(p)
  \quad\text{and}\quad \pis_{1,2} \sim O(p^2)\,.
\end{align}

Let us begin with the $d$ quarks. To use HDET we momentarily switch to
Minkowski space-time. The fermionic part of the microscopic Lagrangian
is then given by
\begin{align}
  \label{eq:L_micr}
  \mathcal{L} = \bar d \big( i \gamma^\nu D_\nu + (\mui + \dMu_d)
  \gamma^0 \big) d - \bar{d}_L M_d d_R - \bar{d}_R M_d^\dagger d_L\,.
\end{align}
In the regime $\mui\gg\Lambda_{\qcd}$ this theory can be treated in
the framework of HDET, where we expand in powers of $\dmu_{d,f}$ in a
way analogous to \cite{Bedaque:2001je} (where QCD at high baryon
density, rather than high isospin density, was considered). To second
order in $p$ the result is then given by
\begin{multline}
  \label{eq:L_hdet}
  \mathcal{L}_\text{HDET} = \sum_{\vec{v}_F} d^\dagger_{R+}(\vec{v}_F) \left( 
  i \check v^\nu D_\nu + \dMu_d -\frac{1}{2\mui}\big( (\slashed{D}_\perp)^2 +
  M_d^\dagger M_d \big)\right) d_{R+}(\vec{v}_F)
  \\
  + \sum_{\vec{v}_F} d^\dagger_{L+}(\vec{v}_F) \left( 
  i \check v^\nu D_\nu + \dMu_d -\frac{1}{2\mui}\big( (\slashed{D}_\perp)^2 + M_d
  M_d^\dagger \big)\right) d_{L+}(\vec{v}_F) + \dots\,,
\end{multline}
where $\check v^\nu=(1,\vec{v}_F)$ with Fermi velocity $\vec{v}_F$,
$D$ and $\slashed{D}_\perp$ are counted as $O(p)$, and the dots denote
higher orders in $p$.  The definitions of the projected modes
$d_{i+}\,(i=R,L)$ and of $\slashed{D}_\perp$ are given in
\cite{Hong:1998tn,Hong:1999ru}.  The first two terms in parentheses
are $O(p)$, while the next two terms are $O(p^2/\mui)$, i.e., the
expansion parameter is $p/\mui$.

In \eqref{eq:L_micr} both $\dMu_d$ and $i\del_0$ come with $\gamma^0$.
Furthermore, in \eqref{eq:L_hdet} the mass matrix and $\dMu_d$ appear
in the combination $\dMu_d - M_d^\dagger M_d/2\mui$ for $d_{R+}$ and
$\dMu_d -M_dM_d^\dagger/2\mui$ for $d_{L+}$.  This implies that
$\mathcal{L}_\text{HDET}$ at this order would be invariant under a
time-dependent $\U(\Nfhalf)_{d_R}\times\U(\Nfhalf)_{d_L}$ flavor
transformation if both $\dMu_d - M_d^\dagger M_d/2\mui$ and $\dMu_d -
M_d M_d^\dagger/2\mui$ transformed as time components of local gauge
fields coupled to $\U(\Nfhalf)_{d_R}\times\U(\Nfhalf)_{d_L}$
\cite{Bedaque:2001je}.  Since according to \eqref{eq:MUV} the NG
fields $U$ and $V$ transform in the $d$-quark sector as
\begin{equation}
  U \to g_{d_L}U \quad\text{and}\quad V\to Vg^\dagger_{d_R} \quad
  \text{for} \quad g_i\in\U(\Nfhalf)_i
\end{equation}
the effective theory can also be made invariant under the spurious
symmetry via the replacements\footnote{That the specific combinations
  of $M_d$ and $\dMu_d$ occurring in \eqref{eq:d0UV} are reasonable
  can be inferred intuitively, i.e., from the Fermi level of a free
  $d$ quark, $p_F$\,. With an insertion of $m_d\ne 0$ and a small
  shift $\mui\to\mui+\dmu_d$ satisfying $m_d, \,\dmu_d\ll \mui$ we
  obtain $p_F =\big((\mui+\dmu_d)^2-m_d^2\big)^{1/2} \simeq
  \mui+\dmu_d-m_d^2/2\mui$, and thus it is the combination
  $\dmu_d-m_d^2/2\mui$ that effectively parameterizes the shift of the
  Fermi level.  For $u$ quarks we need to flip the sign of $\mui$ and
  obtain $\dmu_u+m_u^2/2\mui$.}
\begin{subequations}
  \label{eq:d0UV}
  \begin{align}
    \del_0 U & \to \del_0 U - i \left( \dMu_d -\frac{1}{2\mui}M_d
      M_d^\dagger \right)U \,,\\
    \del_0 V^\dagger & \to \del_0 V^\dagger -i\left( \dMu_d
      -\frac{1}{2\mui}M_d^\dagger M_d \right)V^\dagger\,. 
  \end{align}
\end{subequations}
Note that the second term in parentheses is suppressed by $O(p/\mui)$
with respect to the first term.  Therefore it can be dropped when we
construct the effective Lagrangian to $O(p^2)$.

The $u$-quark sector can be treated in a similar manner. In the end,
after analytic continuation to Euclidean space-time $\del_0\to
i\del_4$, we find the leading $O(p^2)$ effective Lagrangian including
the effects of $\dMu_{u,d}$ to be given by
\begin{align}
  \label{eq:L_eff_full}
  \mathcal{L}_\text{eff}(U,V) = &\ 
  \frac{F^2}{4}\left\{ 
    \Tr[ \nabla_4 U^\dagger \nabla_4 U + v^2 \del_i U^\dagger\del_i U ] + 
    \Tr[ \nabla'_4 V^\dagger \nabla'_4 V + v^2 \del_i V^\dagger\del_i V ] 
  \right\}
  \notag
  \\
  & - f^2 \left\{(\Tr[U^\dagger \nabla_4 U])^2 
    + \tilde v^2 (\Tr[U^\dagger \del_i U])^2
    + (\Tr[V^\dagger \nabla'_4 V])^2
    + \tilde v^2 (\Tr[V^\dagger \del_i V])^2 \right\} 
  \notag 
  \\
  & + \frac{3N_c}{4\pi^2}\Delta^2 \big\{ \!\Tr[M_uU^\dagger M_d V^\dagger] 
  + \text{c.c.} \big\} 
  - \Phi \, \big\{ \!\Tr[ \pis_1 U + \pis_2 V^\dagger ] + 
  \text{c.c.} \big\} \,, 
\end{align}
where\footnote{Note that $\nabla_4 U^\dagger\ne(\nabla_4 U)^\dagger$
  and $\nabla_4' V^\dagger\ne(\nabla_4' V)^\dagger$.}
\begin{subequations}
  \label{eq:UV}
  \begin{align}
  \label{eq:U1}
  \nabla_4 U & = \del_4 U - \dMu_d U + U \dMu_u \,,
  \\
  \label{eq:U2}
  \nabla_4 U^\dagger & = \del_4 U^\dagger + 
  U^\dagger \dMu_d - \dMu_u U^\dagger  \,,
  \\
  \label{eq:V1}
  \nabla'_4 V & = \del_4 V - \dMu_u V + V \dMu_d \,,
  \\
  \label{eq:V2}
  \nabla'_4 V^\dagger & = \del_4 V^\dagger + 
  V^\dagger \dMu_u - \dMu_d V^\dagger \,.
  \end{align}
\end{subequations}
This completes the derivation of the effective theory in the presence
of $\dMu_{u,d}$. The low-energy constants in \eqref{eq:L_eff_full} are
the same as those in \eqref{eq:L_eff}.  In particular, they are
defined in the limit $\dMu_u=\dMu_d=0$.  Equation \eqref{eq:L_eff}
follows as a limit of \eqref{eq:L_eff_full} if we set $\dMu_u = \dMu_d
= 0$.

Let us recall that, when $\mathcal L_\text{eff}$ was constructed in
\eqref{eq:L_eff}, terms with a single derivative such as
$\Tr[U^\dagger \del_4 U]$ were dropped as they are total derivatives.
Retaining this term and replacing $\del_4 U$ by $\nabla_4 U$ according
to \eqref{eq:U1} would result in a non-derivative term,
\begin{align}
  \Tr[U^\dagger \nabla_4 U] = 
  \Tr[U^\dagger \del_4 U] + \Tr[\dMu_u-\dMu_d]\,.
\end{align}
The second term vanishes thanks to \eqref{eq:GF}, so omission of the
single-derivative terms in \eqref{eq:L_eff} does not influence our
current discussion.  In appendix~\ref{app:red} we discuss an ambiguity
that appears if the condition \eqref{eq:GF} is not respected.

If we set $\dMu_u = \dMu_d = \mu_q \1_{\Nfhalf}$, representing a small
common quark chemical potential on top of a large isospin chemical
potential, we find that $\mu_q$ disappears from the covariant
derivatives in \eqref{eq:UV}, leaving no effect on $\mathcal
L_\text{eff}$.  This is the high-isospin-density analogue of the
Silver Blaze phenomenon we mentioned in the introduction. It could
have been anticipated from the fact that the NG modes $\sim \bar u d$
in this theory carry no net baryon number. We expect a nonzero baryon
number to emerge only if $\mu_q$ is greater than $\mu_q^c\sim
\Delta/\sqrt{2}$, at which the isotropic BCS phase gives place to a
new phase, but this is beyond the domain of validity of our low-energy
effective theory.

\subsection[\texorpdfstring{$\epsilon$-expansion}{epsilon-expansion}]{\boldmath
  $\epsilon$-expansion}
\label{sec:eps}

We now move on to the $\epsilon$-regime
\cite{Gasser:1987ah,Leutwyler:1992yt,Yamamoto:2009ey}.  We consider
the system to be confined in a 4-dimensional Euclidean box with linear
extent $L$ and volume $V_4= L^4$ satisfying
\begin{align}
  \frac{1}{\Delta} \ll L \ll \frac{1}{m_\text{NG}}\,,
\end{align}
where $m_\text{NG}$ is the mass scale of the NG fields.  The first
inequality ensures that the contribution of non-NG modes to the
partition function is negligible, while the second inequality implies
that the Compton wavelength of the NG fields is much larger than the
size of the box. In this limit the partition function is dominated by
the zero-momentum modes of the NG fields. This regime can be defined
through the ``$\epsilon$-expansion'' counting\footnote{This should not
  be confused with the conventional $\epsilon$-regime at zero density,
  where $M_{u,d} \sim O(\epsilon^4)$.}
\begin{align}
  \label{eq:e-expansion}
  \del_\nu \sim 1/L \sim \xi(x) \sim O(\epsilon)\,, \quad
  M_{u,d} \sim \dMu_{u,d} \sim O(\epsilon^2)\,,  \quad\text{and}\quad
  \pis_{1,2}\sim O(\epsilon^4)\,.
\end{align}
Here, $\xi(x)$ represents the nonzero-momentum modes of $U$ and $V$,
which are given by $U(x)= U_0 \exp(i\sqrt{2}\xi_U(x)/F)$ and $V(x)=
V_0 \exp(i\sqrt{2}\xi_V(x)/F)$, where $U_0$ and $V_0$ denote the
zero-momentum modes.

Extracting the leading terms up to $O(\epsilon^4)$ from
\eqref{eq:L_eff_full} and discarding higher-order terms we obtain
\begin{align}
  \mathcal L_\text{eff}\big|_{\epsilon^4} = \ 
  & \frac{1}{2}\Tr\!\left[(\del_4\xi_U)^2+v^2(\del_i\xi_U)^2
  + (\del_4\xi_V)^2+v^2(\del_i\xi_V)^2\right]
  \notag 
  \\
  &+ 2 \frac{f^2}{F^2} \left[ (\Tr\del_4\xi_U)^2 + 
  \tilde v^2 (\Tr \del_i \xi_U)^2
  + (\Tr \del_4\xi_V)^2 + 
  \tilde v^2 (\Tr \del_i \xi_V)^2 \right]
  \notag \\
  & + \frac{F^2}{4}\Tr\!\big[
    (U_0^\dagger \dMu_d-\dMu_uU_0^\dagger)(-\dMu_dU_0+U_0\dMu_u) 
    + 
    (V_0^\dagger\dMu_u-\dMu_dV_0^\dagger)(-\dMu_uV_0+V_0\dMu_d)
  \big]
  \notag \\
  & + \frac{3N_c}{4\pi^2}\Delta^2 
  \big\{ \!\Tr[M_uU_0^\dagger M_d V_0^\dagger] + \text{c.c.}
  \big\} 
  - \Phi \, \big\{\!\Tr[ \pis_1 U_0 + \pis_2 V_0^\dagger ] + 
  \text{c.c.}
  \big\} \,.  
  \label{eq:L_epsilon}
\end{align}
In deriving \eqref{eq:L_epsilon} we omitted several terms at
$O(\epsilon^4)$ either because they are total derivatives or because
they are proportional to $\Tr(\dMu_u-\dMu_d)$, which vanishes
according to condition \eqref{eq:GF}.

In the $\epsilon$-regime, the zero-momentum modes are no longer
suppressed as $V_4\to\infty$, and one has to sum up their
contributions nonperturbatively \cite{Gasser:1987ah}. This is in
contrast to the $p$-regime \eqref{eq:p-rule}, where they are counted
as $O(p)$ like nonzero-momentum modes and can be treated
perturbatively.  The kinetic terms for $\xi_U(x)$ and $\xi_V(x)$ in
\eqref{eq:L_epsilon} only affect the multiplicative normalization of
the partition function and are irrelevant for the dependence of the
partition function on $\dMu_{u,d}$, $M_{u,d}$, and $\pis_{1,2}$.
   
We thus find that the finite-volume partition function of QCD for
$\mui\gg \Lambda_{\qcd}$ at leading order of the new
$\epsilon$-expansion \eqref{eq:e-expansion} is given by
\begin{align}
  \label{eq:Z_qcd_ep}
  Z_{\qcd}^{(\Nf)}(\mui; \dMu_{u,d}, M_{u,d}, \pis_{1,2}) = \quad
  \int\limits_{\mathclap{\U(\Nfhalf)}} dU_0  \;
  \int\limits_{\mathclap{\U(\Nfhalf)}} dV_0 \,
  \exp\left( \fbox{A}+\fbox{B}+\fbox{C} \right)
\end{align}
with
\begin{subequations}
  \begin{align}
    \fbox{A} & = 
    -\frac{1}{2}V_4F^2 \Tr\big[ 
    \dMu_u U_0^\dagger \dMu_d U_0 + \dMu_d V_0^\dagger \dMu_u V_0
    -\dMu_u^2-\dMu_d^2 
    \big]  \,,
    \\
    \fbox{B} & = - \frac{3N_c}{4\pi^2}V_4\Delta^2 
    \left\{ \Tr[M_uU_0^\dagger M_d V_0^\dagger] + \text{c.c.} 
    \right\} ,
    \\
    \fbox{C} & = 
    V_4 \Phi \left\{\Tr[ \pis_1 U_0 + \pis_2 V_0^\dagger ] + 
    \text{c.c.}
    \right\} .
  \end{align}
\end{subequations}
This completes the derivation of the effective partition function in
the $\epsilon$-regime. In section~\ref{sec:rmt} we will show that the
expression \eqref{eq:Z_qcd_ep} can be reproduced by a certain
zero-dimensional random matrix theory.  Note that for
$\pis_1=\pis_2=0$, \eqref{eq:Z_qcd_ep} can be computed analytically in
two limits: If at least one of $\dMu_u$ or $\dMu_d$ is zero we obtain
the Berezin-Karpelevich integral \cite{Berezin:1958,Guhr:1996vx}. If
at least one of $M_u$ or $M_d$ is zero we obtain the
Harish-Chandra--Itzykson-Zuber integral
\cite{Harish-Chandra:1958,Itzykson-Zuber:1980}.

\subsection{Sign problem}
\label{sc:signprob}

Consider $N_f=4$ QCD with $\dMu_u=\dMu_d=\mu_q \1_2$,
$\pis_1=\pis_2=0$, and equal mass $m$, i.e.,
\begin{align}
  Z^{(4)}_{\qcd}(\mui; \mu_q, m) 
  & = \Big\langle 
  {\det}^2\big(D(-\mui+\mu_q)+m\big) 
  {\det}^2\big(D(\mui+\mu_q)+m\big) 
  \Big\rangle_{\YM}\,. 
  \label{eq:Z4full}
\end{align}
This theory suffers from a sign problem at $\mu_q\ne 0$. Let us denote
the complex phase of the fermion determinants inside
$\langle\dots\rangle$ by $\ee^{i\theta}$.  To estimate the severity of
the sign problem it is useful to compare the partition function
\eqref{eq:Z4full} with the phase-quenched (\pq) theory,
\begin{align}
  Z^{(4)}_{\rm{\pq}\qcd}(\mui; \mu_q, m) 
  & = \Big\langle 
  \left|{\det}\big(D(-\mui+\mu_q)+m\big)\right|^2 
  \left|{\det}\big(D(\mui+\mu_q)+m\big)\right|^2 
  \Big\rangle_{\YM}
  \notag\\
  & = \Big\langle 
  {\det}\big(D(-\mui+\mu_q)+m\big) 
  {\det}\big(D(\mui-\mu_q)+m\big)
  \notag
  \\
  & \qquad \times 
  {\det}\big(D(\mui+\mu_q)+m\big) 
  {\det}\big(D(-\mui-\mu_q)+m\big) 
  \Big\rangle_{\YM}
  \,. 
  \label{eq:Z4fullpq}
\end{align}
%%%
\begin{figure}[t]
  \centerline{
    \includegraphics[width=.7\textwidth]{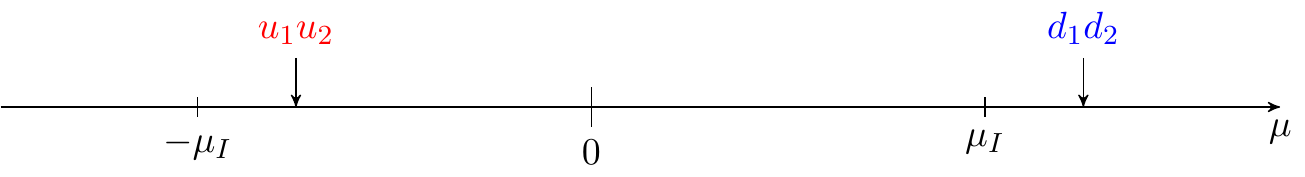}
  }\vspace{\baselineskip}
  \centerline{
    \includegraphics[width=.7\textwidth]{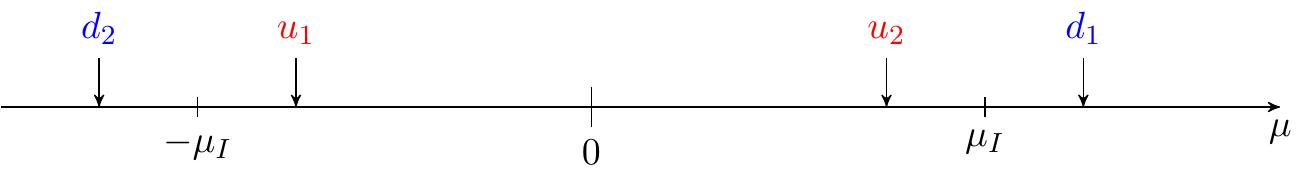}
  }
  \caption{ Chemical potentials of $u$ and $d$ quarks before and after
    phase quenching. The top figure corresponds to
    $Z^{(4)}_{\qcd}(\mui; \mu_q, m)$ in \eqref{eq:Z4full} and the
    bottom figure to $Z^{(4)}_{\rm{\pq}\qcd}(\mui; \mu_q, m)$ in
    \eqref{eq:Z4fullpq}.}
  \label{fg:phasequench} 
\end{figure}
%%% 
The change due to the phase quenching is shown schematically in
figure~\ref{fg:phasequench}.  Then
\begin{align}
  \langle \ee^{i\theta} \rangle_{\rm \pq} & = 
  \frac{Z^{(4)}_{\qcd}(\mui; \mu_q, m)}{Z^{(4)}_{\rm{\pq}\qcd}(\mui; \mu_q, m)}
  \notag\\
  & = \frac{Z^{(4)}_{\qcd}(\mui; \dMu_u=\dMu_d=\mu_q\1_2, m)}
  {Z^{(4)}_{\qcd}(\mui; \dMu_u=\dMu_d=\mu_q \tau_3, m)}\,.
\end{align}
For a rough estimate it suffices to apply the mean-field approximation
by dropping derivative terms in \eqref{eq:L_eff_full}, which leads us
to the microscopic limit \eqref{eq:Z_qcd_ep}. The result, to leading
order in the thermodynamic limit ($V_4\Delta^2m^2\gg 1$ and
$V_4F^2\mu_q^2\gg 1$), is given by\footnote{The integral in
  \eqref{eq:z4} is known exactly
  \cite{Brower:1981vt,Jackson:1996jb,Akuzawa:1998tq} and given by
  $I_0(a)^2-I_1(a)^2$, where $a=3N_cV_4m^2\Delta^2/\pi^2$ and $I_0$
  and $I_1$ are modified Bessel functions.  It would be interesting to
  derive an exact result for the integral in \eqref{eq:z4pq}, but this
  is beyond the scope of the present paper.}
\begin{align}
  Z^{(4)}_{\qcd}(\mui; \dMu_u=\dMu_d=\mu_q\1_2, m)
  & \simeq \int\limits_{\U(2)}\!\!dU \! \int\limits_{\U(2)}\!\!dV 
  \exp\left( - \frac{3N_c}{2\pi^2} V_4 \Delta^2 m^2\mathrm{Re\,Tr}[UV] \right)
  \notag\\
  & \sim ~\exp\left( \frac{3N_c}{\pi^2} V_4 \Delta^2 m^2 \right), 
  \label{eq:z4}
  \\
  Z^{(4)}_{\qcd}(\mui; \dMu_u=\dMu_d=\mu_q \tau_3, m)
  & \simeq \int\limits_{\U(2)}\!\!dU \! \int\limits_{\U(2)}\!\!dV
  \exp\left( - \frac{3N_c}{2\pi^2} V_4 \Delta^2 m^2\mathrm{Re\,Tr}[UV]  + 2V_4F^2\mu_q^2 
  \right. 
  \notag
  \\
  & \left. \hspace*{36mm}
    -\frac{1}{2}V_4F^2\mu_q^2 \Tr[U\tau_3 U^\dagger \tau_3 + V\tau_3 V^\dagger \tau_3]
  \right)
  \notag\\
  &\sim ~ \exp\left(  
    \frac{3N_c}{\pi^2} V_4 \Delta^2 m^2 + 4 V_4F^2\mu_q^2
  \right), 
  \label{eq:z4pq}
\end{align}
where in the last step we have evaluated the integral
for the configuration
\begin{align}
  U = \ee^{i\varphi}(\cos\theta\,\tau_1 + \sin\theta\, \tau_2)
  ~~\text{and}~~
  V = - \ee^{-i\varphi}(\cos\theta\, \tau_1 + \sin\theta\, \tau_2)
\end{align}
for arbitrary $\theta$ and $\varphi$ to maximize the exponent in the
integrand.  Consequently, the sign problem is exponentially hard at
any nonzero $\mu_q$,
\begin{align}
  \langle \ee^{i\theta} \rangle_{\rm \pq} & \sim 
  \ee^{-4 V_4F^2\mu_q^2}\,.
  \label{eq:ssign}
\end{align}
This is in marked contrast to QCD without isospin chemical potential,
where the sign problem becomes severe only for $\mu_q \gtrsim m_\pi/2$
\cite{Splittorff:2006fu,Splittorff:2007ck}. The difference stems from the fact that
$Z^{(4)}_{\rm{\pq}\qcd}(\mui;\mu_q,m)$ contains strictly massless NG
modes that couple to $\mu_q$. Let us recall that at $\mu_q=m=0$ there
were eight NG modes in total. At $m\ne 0$, four of them acquire masses
while the other four remain massless.\footnote{This can be understood
  by looking at the exponent of \eqref{eq:z4}.  The term $\Tr[UV]$
  gives mass to the NG modes.  Only the diagonal subgroups of the two
  coset fields $U$ and $V$ remain massless, which can be seen
  explicitly by substituting $U=\exp(i\pi^a \tau^a)$ (and likewise for
  $V$) and expanding the Lagrangian to second order in the fields.}
Because two of the four massless modes are charged under the $\U(1)$
symmetry to which $\mu_q$ couples in $Z^{(4)}_{\qcd}(\mui;
\dMu_u=\dMu_d=\mu_q \tau_3, m)$, they Bose-condense as soon as a
nonzero $\mu_q$ is turned on. This leads to the exponentially severe
sign problem \eqref{eq:ssign}. By contrast, in QCD without isospin
chemical potential, all pions are gapped for $m\ne 0$. This postpones
the onset of the sign problem until $\mu_q=m_\pi/2$.

The analysis of this subsection can straightforwardly be generalized
to any $N_f$ divisible by 4 since in this case we can flip half of the
chemical potentials and combine the Dirac determinants pairwise to
obtain the absolute value.  However, this is no longer possible for
$N_f\equiv 2\mod 4$.

\section{Random matrix theory and spectral properties}
\label{sec:rmt}

\subsection{Random matrix model for nonzero stress}

A random matrix model that exactly reproduces part $\fbox{B}$ of
\eqref{eq:Z_qcd_ep} has been constructed in \cite{Kanazawa:PhD}. Here
we present an extension of this model to incorporate the effects of
$\dMu_{u,d}$ and $\pis_{1,2}$,
\begin{multline}
  \label{eq:Z_rmt}
  Z^{(\Nf)}_{\rmt} (\Muu_{u,d}, \Mm_{u,d}, \Pis_{1,2}) = 
  \\
  \int\limits_{\mathclap{\mathbbm{C}^{N\times N}}} dP \:
  \int\limits_{\mathclap{\mathbbm{C}^{N\times N}}} dQ \,
  \ee^{-N\Tr(PP^\dagger + QQ^\dagger)} \,
  \det 
  \left(\begin{array}{cc|cc} 
    \Mm_u^\dagger & P - \Muu_u & \Pis_1 & 0
    \\ 
    - Q^\dagger - \Muu_u & \Mm_u & 0 & \Pis_2
    \\\hline 
    \Pis_2^\dagger & 0 & \Mm_d^\dagger & Q - \Muu_d  
    \\ 
    0 & \Pis_1^\dagger & - P^\dagger - \Muu_d & \Mm_d 
  \end{array}\right) ,
\end{multline}
where $P$ and $Q$ are $N\times N$ complex matrices while $\Muu_u$,
$\Muu_d$, $\Mm_u$, and $\Mm_d$ are $(\Nfhalf)\times(\Nfhalf)$ matrices
acting on flavor indices (i.e., we write $P-\Muu_u$ instead of
\mbox{$P\otimes \1_{\Nfhalf} - \1_N\otimes \Muu_u$} etc.\ for
brevity).  All dimensionless parameters carry a hat to distinguish
them from physical variables.\footnote{Note that the RMT quantity
  $\Muu$ corresponds to the physical quantity $\dMu$ and not to
  $\Mu$.} The inclusion of the chemical potentials in this form was
motivated by Stephanov's model \cite{Stephanov:1996ki}, which was
devised for QCD at low baryon density. There is another well-known way
of incorporating the chemical potential into RMT devised by Osborn
\cite{Osborn:2004rf} where the chemical potential is multiplied by
another Gaussian random matrix. We expect such a formulation to belong
to the same large-$N$ universality class as \eqref{eq:Z_rmt}.

As we will show shortly, our model \eqref{eq:Z_rmt} describes QCD at
large isospin chemical potential.  Models with a similar structure
were investigated in \cite{Splittorff:2003cu,Akemann:2004dr} with the
aim of describing QCD at small isospin chemical potential (called
phase-quenched QCD by those authors). These models must not be
confused with ours.  It is worthwhile to note that
refs.~\cite{Splittorff:2003cu,Akemann:2004dr} confirmed through
explicit calculation that the two formulations of incorporating $\mu$
into RMT lead to an identical quenched microscopic spectral
density. This is strong evidence that these two formulations are
indeed equivalent in the large-$N$ microscopic limit.

Let us return to the model \eqref{eq:Z_rmt}. Using standard techniques
(see, e.g., \cite{Shuryak:1992pi,Akemann:2004nw,Kanazawa:PhD}) of
fermionization and Hubbard-Stratonovich transformation, we find that
in the large-$N$ limit with the scaling
\begin{align}
  \Mm_{u,d}\sim \Muu_{u,d} \sim O(1/\sqrt{N}) \quad\text{and}\quad
  \Pis_{1,2} \sim O(1/N)\,,
\end{align}
\eqref{eq:Z_rmt} reduces to a nonlinear sigma model, 
\begin{multline}
  \label{eq:Z_rmt_sigma}
  Z^{(\Nf)}_{\rmt} (\Muu_{u,d},\Mm_{u,d},\Pis_{1,2}) = \quad
  \int\limits_{\mathclap{\U(\Nfhalf)}} dU \;\;
  \int\limits_{\mathclap{\U(\Nfhalf)}} dV \,
  \exp\Big( N\Tr\Big[
  - \Muu_u U^\dagger \Muu_d U - \Muu_d V^\dagger \Muu_u V 
  \\
  + ( - \Mm_u U^\dagger \Mm_d V^\dagger + \text{c.c.} )  
  + ( \Pis_1 U + \Pis_2 V^\dagger + \text{c.c.} ) 
  \Big] \Big)\,.
\end{multline}
Comparing \eqref{eq:Z_rmt_sigma} with \eqref{eq:Z_qcd_ep} we find the
correspondence
\begin{align}
  \label{eq:Z_Z_equal}
  Z_{\qcd}^{(\Nf)}(\mui; \dMu_{u,d}, M_{u,d}, \pis_{1,2}) =  
  \ee^{ N \Tr(\Muu_u^2+\Muu_d^2) }
  Z^{(\Nf)}_{\rmt} (\Muu_{u,d},\Mm_{u,d},\Pis_{1,2})
\end{align}
with the identifications 
\begin{subequations}
  \label{eq:match}
  \begin{align}
    \sqrt{\frac{V_4F^2}{2}} \, \dMu_{u,d} 
    & \Longleftrightarrow \sqrt{N} \Muu_{u,d}\, ,
    \\
    \sqrt{\frac{3N_c}{4\pi^2}V_4\Delta^2} \, M_{u,d} 
    & \Longleftrightarrow \sqrt{N} \Mm_{u,d} \, , 
    \\
    V_4 \Phi \pis_{1,2} & \Longleftrightarrow N \Pis_{1,2} \, .
  \end{align}
\end{subequations}
This proves the equivalence of the partition function for low-energy
QCD in the $\epsilon$-regime and chiral RMT, both at large $\mui$.
Let us add a few comments.
\begin{enumerate}
\item Just as the quark mass couples to the Dirac eigenvalues, the
  pionic source \eqref{eq:sce} couples to the singular values of the
  Dirac operator \cite{Kanazawa:2011tt}. Therefore the above
  correspondence, including the $\Pis_{1,2}$ terms, shows not only the
  equivalence between QCD and RMT for the Dirac eigenvalue
  distribution, but also for the singular-value distribution of the
  Dirac operator.  While a complete proof would necessitate partially
  quenched ChPT \cite{Osborn:1998qb,Basile:2007ki}, in this paper we
  shall be satisfied with the equivalence at the level of the
  fermionic partition function.
\item Within RMT there is no parameter corresponding to $\mui$.  The
  effect of $\mui$ is included implicitly in $\Delta$, $F$, and $\Phi$
  in \eqref{eq:match}. This is true in RMT for two-color QCD at high
  baryon density as well \cite{Kanazawa:2009en,Akemann:2010tv}.
\item The two partition functions in \eqref{eq:Z_Z_equal} differ by a
  factor $\ee^{N \Tr(\Muu_u^2+\Muu_d^2)}$. This factor does not affect
  expectation values in both theories and is irrelevant, unless one is
  interested in the partition function itself, or in its derivative
  w.r.t.\ $\Muu_{u,d}$. Actually such a discrepancy generally arises
  when matching QCD and chiral RMT with chemical potential
  \cite{Akemann:2007rf,Ipsen:2012ug}.
\item It was shown in \cite{Kanazawa:PhD} that $Z^{(\Nf)}_{\rmt}
  (\Mm_{u,d})$ for $\Muu_{u,d}=\Pis_{1,2}=0$ may be cast into the form
  of the determinant of a certain matrix of dimension $\Nfhalf$. Using
  this result one can derive the microscopic spectral density of the
  Dirac matrix $\displaystyle \begin{pmatrix} 0 & P \\
    -Q^\dagger & 0 \end{pmatrix}$ analytically for arbitrary
  masses \cite{Kanazawa:PhD}.  However, such a simple formula is not
  known for the current extension to nonzero $\Muu_{u,d}$.
\end{enumerate}
In order to study the spectral properties of the Dirac matrices 
\begin{align}
  \begin{pmatrix} 0 & P - \Muu_u \\ - Q^\dagger - \Muu_u &
    0 \end{pmatrix} 
  \quad\text{and}\quad
  \begin{pmatrix} 0 & Q - \Muu_d \\ - P^\dagger - \Muu_d &
    0 \end{pmatrix}
\end{align}
the first step would be to derive the eigenvalue representation of the
partition function \eqref{eq:Z_rmt}. However, this is a difficult task
even in the limit $\Mm_{u,d}=\Pis_{1,2}=0$, and we postpone this to
future work. In the next section we turn to the singular values of
these matrices to discuss the Silver Blaze phenomenon of QCD at high
isospin density, where we will find that a number of insights can be
gained without any additional calculation of spectral correlations.

\subsection{Mapping high isospin to low baryon density}
\label{sec:silver}

In the following we set $\Muu_u=\Muu_d=\muu_q\1_{\Nfhalf}$ for simplicity,
which satisfies condition \eqref{eq:GF}.  From the mapping between RMT
and QCD in the $\epsilon$-regime found in the previous subsection we
have the exact correspondence 
\begin{align}
    \label{eq:dirmat}
  \begin{array}{@{\quad}c@{\quad}c@{\quad}c@{\quad}}
    \hline \hline 
    \qcd~(\mui\gg \Lambda_{\qcd}) & & \rmt
    \\\hline
    D(-\mui+\mu_q) & \Longleftrightarrow & 
    \begin{pmatrix}
      0 & P - \muu_q \\ - Q^\dagger - \muu_q & 0 
    \end{pmatrix}^{\mathstrut}
    \\
    D(\mui+\mu_q) & \Longleftrightarrow & 
    \begin{pmatrix}
      0 & Q - \muu_q \\ - P^\dagger - \muu_q & 0 
    \end{pmatrix}_{\mathstrut}^{\mathstrut}
    \\ \hline \hline 
  \end{array}
\end{align}
As remarked above, it is technically difficult to compute the
eigenvalue correlations of these matrices. However, as will be shown
below, one can analytically compute the eigenvalue correlations for
the \emph{product} of these matrices,
\begin{align}
  \label{eq:dirmat2}
  -D(-\mui+\mu_q)D(\mui+\mu_q) \Longleftrightarrow
  \begin{pmatrix}
    (P - \muu_q)(P^\dagger + \muu_q) & 0 \\ 
    0 & (Q^\dagger + \muu_q)(Q - \muu_q) 
  \end{pmatrix}.
\end{align} 
Note that for $\mu_q=0$ the operator on the LHS equals
$D(\mui)^\dagger D(\mui)$,%
\footnote{This follows from $D(-\mu)= - D(\mu)^\dagger$.}  whose
eigenvalues $\{\xi_n^2\}$ are real and nonnegative.  Their positive
square roots $\{\xi_n\}$ (with $\xi_n\ge0$ for all $n$) are called the
singular values of $D(\mui)$.  As a generalization, we will refer to
the positive \emph{and} negative square roots of the eigenvalues of
the operator in \eqref{eq:dirmat2} as the ``stressed singular
values''.  They are no longer real for $\mu_q\ne0$. In the limit
$\mu_q\to 0$ they reduce to $\{+\xi_n\}\cup\{-\xi_n\}$, i.e., the
singular values of $D(\mui)$ and their negatives.  We will show in
section \ref{sc:picond} that the stressed-singular-value spectrum
encodes essential information on the pionic condensate $\langle\bar
u\gamma_5 d\rangle$.

In the remainder of this section 
we concentrate on the influence of nonzero $\muu_q$ by
setting $\Mm_{u,d}=0$.\footnote{The quark-mass dependence of the Dirac
  spectrum at high isospin density was investigated in
  \cite{Kanazawa:PhD}.} Furthermore we assume
$\Pis_1=\piss_1\1_{\Nfhalf}$ and $\Pis_2=\piss_2\1_{\Nfhalf}$ from now
on. Then the partition function \eqref{eq:Z_rmt} reads
\begin{align}
  & Z^{(\Nf)}_{\rmt} (\muu_{q}, 0, \Pis_{1,2})  
  \notag
  \\
  & = 
  \iint dP\, dQ \,
  \ee^{-N\Tr(PP^\dagger + QQ^\dagger)} \,
  {\det}^{\Nfhalf} \begin{pmatrix}
    \piss_1 & P - \muu_q 
    \notag\\
    -P^\dagger - \muu_q & \piss^*_1 
  \end{pmatrix}
  {\det}^{\Nfhalf} \begin{pmatrix}
    \piss_2^* & Q - \muu_q 
    \\
    -Q^\dagger - \muu_q & \piss_2 
  \end{pmatrix}
  \\
  & = 
  \int dP\,\ee^{-N\Tr PP^\dagger} 
  \prod_{k=1}^{N} \big( \piss_1\piss_1^* + p_k^2 \big)^{\Nfhalf}
  \int dQ\, \ee^{-N\Tr QQ^\dagger} 
  \prod_{\ell=1}^{N} \big( \piss_2\piss_2^* + q_\ell^2 \big)^{\Nfhalf}  \,,
  \label{eq:Z_mu_q}
\end{align}
where the $\{\pm i p_k\}$ and $\{\pm i q_\ell\}$ are the eigenvalues ($2N$
each) of
\begin{align}
  \begin{pmatrix} 0 & P - \muu_q 
    \\ - P^\dagger - \muu_q  & 0 \end{pmatrix}
  \quad\text{and}\quad
  \begin{pmatrix} 0 & Q - \muu_q 
  \\ - Q^\dagger - \muu_q  & 0 \end{pmatrix},
\end{align}
respectively.  Since $\muu_q$ enters the Dirac matrices as an
anti-Hermiticity-breaking parameter, the spectra $\{\pm i p_k\}$ and
$\{\pm i q_\ell\}$ spread from the imaginary axis to the entire
complex plane, marking the emergence of the sign problem for the
weight \eqref{eq:Z_mu_q}.  Note that by definition the set $\{\pm
p_k\}\cup\{\pm q_k\}$ constitutes the stressed singular values of the
Dirac operator.

We now notice an interesting fact: the measure in \eqref{eq:Z_mu_q}
consists of two components, each of which is mathematically identical
to the massive partition function of RMT for QCD with $\Nfhalf$
flavors at small quark chemical potential and vanishing isospin
chemical potential \cite{Stephanov:1996ki,Osborn:2004rf},
\begin{align}
  Z^{(\Nfhalf)}_{\rmt} (\muu_q, \mm)_{\nu=0} &= \quad
  \int\limits_{\mathclap{\mathbb{C}^{N\times N}}}
  dP\,\ee^{-N\Tr PP^\dagger}{\det}^{\Nfhalf}\begin{pmatrix}
    \mm^* & P - \muu_q \\ -P^\dagger - \muu_q & \mm 
  \end{pmatrix}
  \label{eq:Z_zero}
  \\
  & = \quad \int\limits_{\mathclap{\mathbb{C}^{N\times N}}}
  dP\,\ee^{-N\Tr PP^\dagger} 
  \prod_{k=1}^{N} \big( \mm\mm^* + p_k^2 \big)^{\Nfhalf}\,, 
  \notag
\end{align}
where the subscript $\nu=0$ implies the restriction to the
topologically trivial sector. According to this exact correspondence,
the universal microscopic correlation functions for the stressed
singular values $\{\pm p_k\}$ and $\{\pm q_\ell\}$ with weight
\eqref{eq:Z_mu_q} are precisely given by those of the well-known
matrix model \eqref{eq:Z_zero}, provided that the pionic sources 
$\piss_1$ and $\piss_2$ in \eqref{eq:Z_mu_q} are identified with 
the quark masses $\mm$ in \eqref{eq:Z_zero}.  
The microscopic correlation functions in the model \eqref{eq:Z_zero}  
have been computed exactly in \cite{Osborn:2004rf} using orthogonal 
polynomials and in \cite{Splittorff:2003cu,Akemann:2004dr} from 
the replica limit of the Toda lattice equation.

In the following three subsections we present insights that can be
gained from earlier works through the mapping from \eqref{eq:Z_mu_q}
to \eqref{eq:Z_zero}.

\subsection{Microscopic stressed-singular-value spectrum}

\begin{figure}[t]
  \centerline{\includegraphics[width=.9\textwidth]{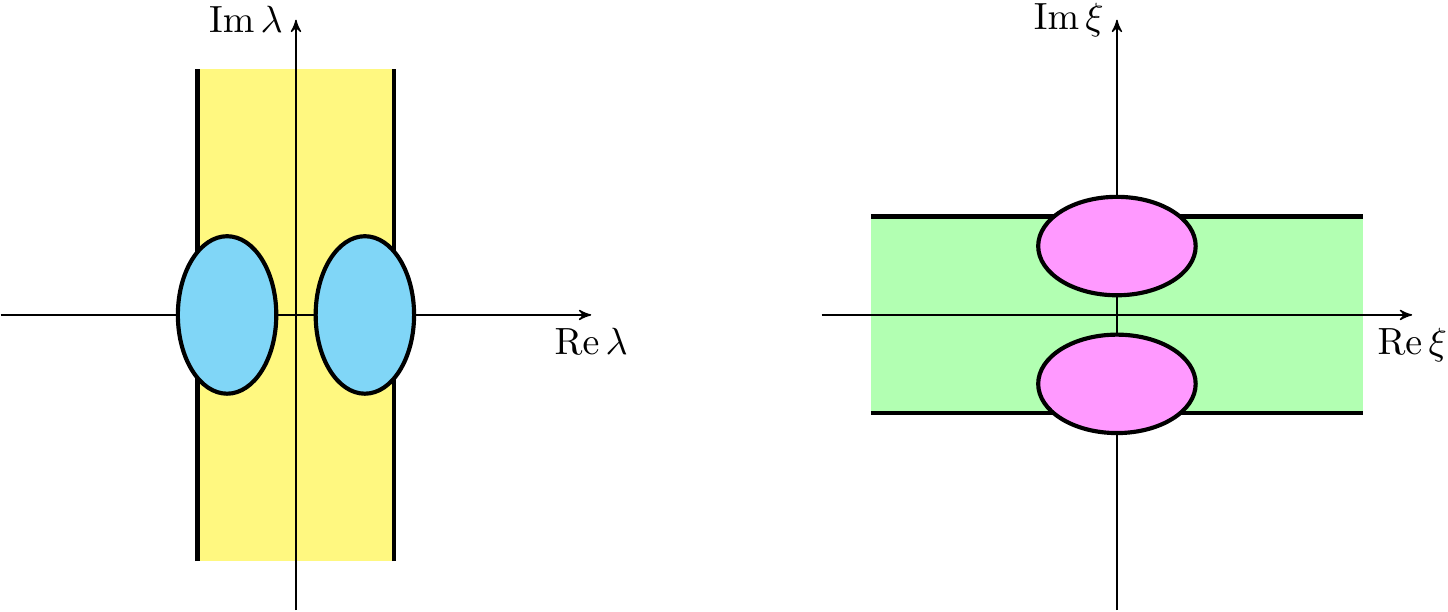}}
  \caption{Left: Sketch of the Dirac spectral density
    $\rho_D(\mu_q;\lambda)$ defined in \eqref{eq:rhoD} for QCD at
    small quark chemical potential $\mu_q>m_\pi/2$.  It is roughly
    constant in the yellow region and strongly oscillating in the blue
    elliptical regions (whose boundaries have been computed in
    \cite{Osborn:2008ab}).  Right: Sketch of the
    stressed-singular-value density $\rho_\text{sv}(\mui,\mu_q;\xi)$
    defined in \eqref{eq:rhosv} for $\mui\gg\Lambda_\text{QCD}$ and
    $\mu_q>\sqrt{\Phi\pis_{1,2}}/F$.  It behaves just like
    $\rho_D(\mu_q;\lambda)$, except that the real and imaginary parts
    are interchanged.}
  \label{fg:stresseddensity}
\end{figure}

It is well known that the microscopic spectral density of the Dirac
operator in QCD with $\mu_q^2\ll 1/\sqrt{V_4}$ and $\mui=0$ changes
its behavior qualitatively as a function of $\mu_q$
\cite{Akemann:2004dr,Osborn:2005ss,Osborn:2008ab}.  At $\mu_q=0$ the
spectral density is supported only on the imaginary axis, and its
value at the origin is proportional to $\Sigma_0$ in the chiral limit,
as known from the Banks-Casher relation \cite{Banks:1979yr}. (Here,
$\Sigma_0$, $F_0$, etc.\ denote low-energy constants of ChPT in the
QCD vacuum.)  For $0<\mu_q<m_{\pi}/2$ the spectral density is roughly
constant on a two-dimensional straight band along the imaginary axis,
with\footnote{We include $1/V_4$ in the definition of the spectral
  density, see \eqref{eq:rhoD} below.}
\begin{align}
  \text{width} \sim \frac{F_0^2\mu_q^2}{\Sigma_0} \quad\text{and}\quad
  \text{height} \sim \frac{\Sigma_0^2}{F_0^2\mu_q^2}
  \qquad (\mui=0)\,.
\end{align}
As $\mu_q$ exceeds $m_{\pi}/2$, the spectral density develops an
elliptical domain of strong oscillations, with an amplitude that
scales exponentially with $V_4$ and a period that shrinks as $1/V_4$
(see figure~\ref{fg:stresseddensity} left).  The spectral density is
no longer real and positive, signaling the onset of a severe sign
problem for $\mu_q>m_\pi/2$.

Through the mapping explained in section~\ref{sec:silver}, these
mathematical results carry over to the regime with
$\mui\gg\Lambda_{\qcd}$ and $\mu_q\ne 0$.  For a physical
interpretation of the mathematical formulas we need to (i) trade the
quark masses for the pionic sources $\pis_{1,2}$, (ii) set the number
of flavors to $\Nfhalf$, and (iii) replace the low-energy constants in
the QCD vacuum by those in the high-isospin-density chiral effective
theory \eqref{eq:L_eff_full}.  In particular, the chiral condensate is
mapped to the pionic condensate.

Instead of quoting complicated mathematical formulas from earlier
works, we would like to discuss the overall structure of the
stressed-singular-value spectrum. For $\mui\gg\Lambda_{\qcd}$ and
$\mu_q=0$, the square roots of the eigenvalues
of the operator $-D(-\mui+\mu_q)D(\mui+\mu_q)\overset{\mu_q=0}{=}$
$D(\mui)^\dagger D(\mui)$ are the singular values of the Dirac operator $D(\mui)$,
as explained after \eqref{eq:dirmat2}. The associated
Banks-Casher-type and Smilga-Stern-type relations have been derived in
\cite{Kanazawa:2011tt,Kanazawa:2012zr}.  For $0<\mu_q
\lesssim\sqrt{\Phi\pis_{1,2}}/F\sim
\sqrt{\pis_{1,2}\Delta/g}\,$,\footnote{Here we used the relations
  $F\sim\mui$ and $\Phi\sim \mui^2\Delta/g$ valid at asymptotically
  high density \cite{Alford:2007xm}.} the (positive and negative)
square roots of the eigenvalues of $-D(-\mui+\mu_q)D(\mui+\mu_q)$,
i.e., the stressed singular values, extend to the two-dimensional
complex plane and have a density that is roughly constant over a
straight band along the real axis, with
\begin{align}
  \text{width} \sim \frac{F^2\mu_q^2}{\Phi}\sim \frac{g\mu_q^2}{\Delta} 
  \quad\text{and}\quad\text{height}
  \sim \frac{\Phi^2}{F^2\mu_q^2} \sim \frac{\mui^2\Delta^2}{g^2\mu_q^2}
  \qquad (\mui\gg\Lambda_{\qcd})\,.
\end{align}
For $\mu_q\gtrsim\sqrt{\pis_{1,2}\Delta/g}$ a severe sign problem sets
in: the flat stressed-singular-value density is invaded by an
elliptical domain of strong oscillations that amplify with $V_4$ as
described above (see figure~\ref{fg:stresseddensity} right).  In
particular, for $\pis_1=\pis_2=0$ the sign problem sets in as soon as
a nonzero $\mu_q$ is turned on, as we have seen in section~\ref{sc:signprob}.

In this manner one can attain a quantitative picture of the
microscopic domain of the operator $-D(-\mui+\mu_q)D(\mui+\mu_q)$ by
simply translating known formulas for $D(\mu_q)$ with $\mui=0$ and
$\Nfhalf$ flavors to the high-isospin-density regime $\mui\to\infty$
with $\Nf$ flavors. It may seem surprising that the same formulas
apply to the description of two seemingly unrelated operators, in two
radically distinct situations.  We can interpret this finding as a
notable manifestation of the universal applicability of RMT.

In the present treatment we have neglected nonzero quark masses.
Understanding their effect on the stressed-singular-value spectrum is
an intriguing problem that is left for future work.

\subsection{Pionic condensate and stressed singular values}
\label{sc:picond}

Let us begin with $\mui=0$. For $\mu_q\ne 0$, the Dirac eigenvalues
spread over the complex plane and the Banks-Casher relation ceases to
be valid, but the Dirac spectral density is still related to the
chiral condensate through the relation 
\begin{align}
  \label{eq:c_c}
  \langle \bar \psi \psi \rangle 
  = \lim_{m\to 0}\lim_{V_4\to\infty}N_f \int_{\mathbb{C}}d\lambda \,
  \frac{2m}{-\lambda^2+m^2}\,\rho_D(\mu_q;\lambda)  
\end{align}
with the Dirac eigenvalue density%
\footnote{The delta function in the complex plane is defined as $\delta(\lambda) \equiv 
\delta(\mathrm{Re}\,\lambda)\delta(\mathrm{Im}\,\lambda)$. }
\begin{align}
  \label{eq:rhoD}
  \rho_D(\mu_q; \lambda)\equiv \frac{1}{V_4}\Big\langle \!\Tr
  \delta\big(\lambda - D(\mu_q)\big) \Big\rangle_{N_f} \,,
\end{align}
where $\Tr\delta(\lambda-A)$ is shorthand for $\sum_i\delta(\lambda -
a_i)$ with $a_i$ the eigenvalues of $A$.  At zero temperature, a
general thermodynamic argument suggests that observables must be
independent of $\mu_q$ for $\mu_q<\mu_q^C \simeq M_N/N_c$, where $M_N$
is the nucleon mass.  This is referred to as the Silver Blaze
phenomenon of dense QCD \cite{Cohen:2003kd,Cohen:2004qp}. Therefore
the chiral condensate \eqref{eq:c_c} must also be independent of
$\mu_q$, despite the fact that $\rho_D(\mu_q; \lambda)$ strongly
varies as a function of $\mu_q$, as illustrated in the last subsection
and in figure~\ref{fg:stresseddensity} left.  This puzzling situation
was investigated mathematically in the microscopic limit
\cite{Osborn:2005ss,Osborn:2008jp}.  The authors found that the
explanation for the $\mu_q$-independent chiral condensate may be
attained through properties of suitable orthogonal polynomials in the
complex plane, which lead to nontrivial cancellations of oscillating
contributions in the integral \eqref{eq:c_c}. They also realized that
it is the whole spectral density, including the flat strip as well as
the strongly oscillating domain, that is responsible for the correct
behavior of $\langle\bar\psi\psi\rangle$ as a function of $m$.

Let us see how these findings add to our understanding of
high-isospin-density QCD. For $\mui\ne 0$, the partition function in
the chiral limit for $\pis_{1}=\pisss\1_{\Nfhalf}$ and
$\pis_{2}=-\pisss\1_{\Nfhalf}$ reads
\begin{align}
  Z_{\qcd}^{(\Nf)}(\mui;\mu_q,\pisss) = 
  \Big\langle 
    {\det}^{\Nfhalf} \big[\! - D(-\mui+\mu_q)D(\mui+\mu_q) + \pisss^2 \big]
  \Big\rangle_{\YM} \,.
\end{align}
It follows from \eqref{eq:sce} that the condensate is 
\begin{align}
  \langle \bar u \gamma_5 d - \bar d \gamma_5 u \rangle & 
  = \lim_{\pisss\to 0}\lim_{V_4\to\infty}\frac{\Nf}{V_4}
  \left\langle \Tr \frac{\omega}{-D(-\mui+\mu_q)D(\mui+\mu_q)+\pisss^2}
  \right\rangle_{\Nf}
  \notag \\
  & =  \lim_{\pisss\to 0}\lim_{V_4\to\infty}\Nf
  \int_{\mathbb{C}}d\xi \,\frac{\pisss}{\xi^2+\pisss^2}\,
  \rho_\text{sv}(\mui,\mu_q; \xi)
  \label{eq:ud_sv}
\end{align}
with the stressed-singular-value density
\begin{align}
  \label{eq:rhosv}
  \rho_\text{sv}(\mui,\mu_q; \xi) \equiv \frac{1}{V_4}
  \Big\langle \sum_n \delta(\xi - \xi_n) \Big\rangle_{\Nf}\,,
\end{align}
where the $\xi_n^2$ are the eigenvalues of
$-D(-\mui+\mu_q)D(\mui+\mu_q)$. A sketch of $\rho_\text{sv}$ is given
in figure~\ref{fg:stresseddensity} right.

As is clear from \eqref{eq:ud_sv} and \eqref{eq:c_c}, the relation
between the stressed-singular-value density
$\rho_\text{sv}(\mui,\mu_q; \xi)$ and the pionic condensate is the
same as the relation between the spectral density $\rho_D(\mu_q;
\lambda)$ and the chiral condensate.  In the microscopic domain, the
two densities are given by the same functions, as noted in the
previous subsection, so all findings for the Dirac spectral density at
$\mu_q\ne 0$ apply to the high-isospin-density regime.  The
$\mu_q$-independence of the pionic condensate (i.e., the \SB{}) at
zero temperature and its discontinuity as $\pisss$ crosses zero can be
accounted for by the same mathematical mechanism as found for the
chiral condensate in \cite{Osborn:2005ss,Osborn:2008jp}. The puzzle
that the $\mu_q$-dependent function $\rho_\text{sv}(\mui,\mu_q; \xi)$
leads to a constant pionic condensate is resolved in this way.

It must be emphasized, though, that the mechanism behind the Silver
Blaze phenomena at low baryon and high isospin density is not the
same. On the one hand, the QCD vacuum does not respond to small
$\mu_q>0$ since it cannot excite a nucleon. On the other hand, dense
isospin matter is insensitive to small $\mu_q>0$ because it is not
energetically preferable to break the Cooper pairs of $\bar u$ and $d$
quarks. It is intriguing that the same mathematical resolution applies
to those two radically different situations.

\subsection{Baryon-number Dirac spectrum}

In this section we discuss the quark-number density $n_q(\mu_q)$,
which is obtained from the partition function as
\begin{align}
  n_q(\mu_q) = \frac{1}{V_4}\frac{d}{d\mu_q} \log Z_{\qcd}(\mu_q)  \,.
\end{align}

We again begin with the low-density regime with $\mui=0$. Expressing
the partition function in terms of the Dirac operator $D(0)$ at zero
chemical potential, we find
\begin{align}
  n_q(\mu_q) & = 
  \frac{1}{V_4}\frac{d}{d\mu_q} \log 
  \Big\langle {\det}^{N_f}  \big(D(0)+m-\mu_q\gamma_4 \big) \Big\rangle_{\YM}
  \notag\\
  &= \frac{1}{V_4}\frac{d}{d\mu_q} \log 
  \Big\langle {\det}^{N_f}  \big(\mu_q-\gamma_4[D(0)+m] \big) \Big\rangle_{\YM}
  \notag\\
  & = \frac{N_f}{V_4} \left\langle \Tr\frac{1}{\mu_q-\gamma_4[D(0)+m]} \right\rangle_{N_f} 
  \notag\\
  & = N_f \int_\mathbb{C}dz \,
  \frac{\rho_q(\mu_q,m; z)}{\mu_q - z}
\end{align}
with
\begin{align}
  \rho_q(\mu_q,m; z)\equiv \frac{1}{V_4}\Big\langle \! \Tr\,
    \delta\big(z - \gamma_4[D(0)+m]\big) \Big\rangle_{N_f}\,. 
  \label{eq:rhoq}
\end{align}
Physically one expects $n_q(\mu_q)$ at $T=0$ to vanish for $0\leq
\mu_q\lesssim M_N/N_c$. This property of QCD was discussed in
connection with the spectral properties of $\gamma_4[D(0)+m]$ in
\cite{Cohen:2003kd,Cohen:2004qp}. Recently this issue was revisited in
\cite{Ipsen:2012ug}, where $\rho_q(\mu_q,m; z)$ was computed
explicitly for $m=0$ in the microscopic limit, i.e., for
$\lambda\sim\mu_q\sim O(1/\sqrt{V_4}F_0)$.

Next we proceed to the regime $\mui\gg \Lambda_{\qcd}$. Since
the condensate $\langle\bar u\gamma_5 d\rangle$ does not carry net
baryon charge, the quark-number density must vanish identically for
$\mu_q$ below a threshold $\sim \Delta/\sqrt{2}$ at which a
phase transition occurs (as reviewed in
section~\ref{sec:intro}).  For simplicity we will only consider
degenerate masses, ignore $\pis_{1,2}$, and set $\dMu_u=\dMu_d=\mu_q
\1_{\Nfhalf}$. From \eqref{eq:Z_gen} we then obtain
\begin{align}
  Z^{(\Nf)}_{\qcd}(\mui,\mu_q,m) = \Big\langle {\det}^{\Nfhalf}\big(\mu_q
  - D_q \big)\Big\rangle_{\YM}
\end{align}
with
\begin{align}
  D_q \equiv 
  \begin{pmatrix}
    \gamma_4[D(-\mui)+m] & 0 \\ 0 & \gamma_4[D(\mui)+m]
  \end{pmatrix} .
\end{align}
Therefore the quark-number density is given by 
\begin{align}
  \label{eq:n_R}
  n_q(\mu_q) = \frac{\Nf}2 \int_{\mathbb{C}}dz \,
  \frac{R_q(\mui,\mu_q; z)}{\mu_q - z} 
\end{align}
with
\begin{align}
  R_q(\mui,\mu_q; z)\equiv 
  \frac{1}{V_4}\Big\langle \!\Tr\, \delta(z - D_q)
  \Big\rangle_{\Nf}\,. 
\end{align}
The spectral density $R_q$ can be computed in the microscopic domain
$\lambda \sim \mu_q \sim O(1/\sqrt{V_4}F)$ using RMT.  From
\eqref{eq:Z_rmt} the corresponding random matrix can be read off as
\begin{align}
  \begin{array}{@{\quad}c@{\quad}c@{\quad}c@{\quad}}
  \hline \hline 
  \qcd~(\mui\gg \Lambda_{\qcd}) & & \rmt
  \\\hline
  \gamma_4[D(-\mui)+m] & \Longleftrightarrow & 
  \begin{pmatrix}
    - Q^\dagger & \mm \\ \mm & P  
  \end{pmatrix}^{\mathstrut}
  \\
  \gamma_4[D(\mui)+m] & \Longleftrightarrow & 
  \begin{pmatrix}
    - P^\dagger & \mm \\ \mm & Q  
  \end{pmatrix}_{\mathstrut}^{\mathstrut}
  \\ \hline \hline 
  \end{array}
\end{align}
It is a challenging task to compute the spectral density of these
random matrices. However, the problem simplifies considerably if we
take the chiral limit $\mm=0$, as $P$ and $Q$ are then decoupled. The
spectral density of the simplified matrices was worked out
analytically in \cite{Ipsen:2012ug} in an effort to find
$\rho_q(\mu_q,m; z)$ at $\mui=0$, cf.~\eqref{eq:rhoq}.  The
mathematical equivalence between \cite{Ipsen:2012ug} and this work
enables us to extract information for $R_q(\mui,\mu_q; z)$ with no
additional calculation.  Adapting the findings of \cite{Ipsen:2012ug}
to our context, we can conclude the following.
\begin{enumerate}
\item At $\mu_q=0$, the sign problem is absent and the density
  $R_q(\mui,0; z)$ is positive definite.  In the macroscopic regime it
  varies smoothly, and in the microscopic regime it is actually
  constant: $R_q(\mui,0; z)\sim F^2\sim \mui^2$.  As $\mu_q$
  increases from zero, a circular domain of radius $\mu_q$ appears
  around the origin in which $R_q(\mui,\mu_q; z)$ shows extremely
  rapid oscillations with amplitude growing exponentially with $V_4$,
  similarly to what is observed in the Dirac spectral density at
  $\mu_q\ne 0$.
\item The quark-number density $n_q(\mu_q)$ follows from
  $R_q(\mui,\mu_q; z)$ via \eqref{eq:n_R}. If the integral is
  computed using only the constant part of $R_q$, the resulting $n_q$
  increases monotonically with $\mu_q$, in apparent contradiction
  with the expected Silver Blaze phenomenon of dense isospin
  matter. However, inclusion of the oscillating part of the spectrum
  cures this problem, and the resulting $n_q$ shows the correct
  $\mu_q$-independence.\footnote{To prove this, the numerical factor
  $\ee^{N\Tr(\Muu_u^2+\Muu_d^2)}$ in \eqref{eq:Z_Z_equal} must 
  be taken into account.}  
\end{enumerate}

\section{Comment on two-color QCD}
\label{sec:2color}

While the main body of this paper concentrates on QCD with $N_c\geq
3$, it seems worthwhile to comment on possible extensions of this work
to two-color QCD, because the finite-density dynamics of the latter
has been actively explored in lattice simulations (see, e.g.,
\cite{Hands:1999md,Kogut:2001na,Hands:2011ye}).  To avoid
complications we will only consider the case $\dMu_u = \dMu_d = \mu_q
\1_{\Nfhalf}$.  First and foremost, the Dirac operator for $\SU(2)$
gauge group possesses an anti-unitary symmetry $C\tau_2\gamma_5
D(\mu)C\tau_2\gamma_5=D(\mu)^*$, with $\tau_2$ the second generator of
$\SU(2)$ \cite{Leutwyler:1992yt}. As a consequence, the partition 
function of two-color QCD is invariant under the exchange of quark chemical 
potential and isospin chemical potential \cite{Splittorff:2000mm}: 
\begin{align}
  Z^{(\Nf)}_{N_c=2}(\mui; \{\dmu\},m) 
  & = \Big\langle 
  {\det}^{\Nfhalf}\big(D(-\mui+\mu_{q})+m \big)\,
  {\det}^{\Nfhalf}\big(D(\mui+\mu_{q})+m \big)
  \Big\rangle_{\YM}
  \notag\\
  & = \Big\langle 
  {\det}^{\Nfhalf}\big(D(\mui-\mu_{q})+m \big)\,
  {\det}^{\Nfhalf}\big(D(\mui+\mu_{q})+m \big)
  \Big\rangle_{\YM}\,. 
  \label{eq:z2color}
\end{align}
The patterns of symmetry breaking with or without chemical potentials 
are summarized in table~\ref{tb:2color1}. It is notable that, unlike in QCD 
with $N_c\geq 3$, the quark chemical potential $\mu_q$ in two-color QCD 
enters as a symmetry-breaking external field.
\begin{table}[t]
  \begin{center}
    \begin{tabular}{|c||c|c|}
      \hline 
      & $N_c=2$ & $N_c\geq 3$ 
      \\\hline\hline 
      $\mui=\mu_q=0$ & $\SU(2N_f)\to \Sp(2N_f)$ & $\SU(N_f)_R\times
      \SU(N_f)_L \to \SU(N_f)_V$ 
      \\\hline 
      $\mui\ne 0$,~$\mu_q= 0$ 
      & $\displaystyle\begin{matrix}
        \U(N_f)_R\times \U(N_f)_L 
        \\
        \to \Sp(N_f)_R\times \Sp(N_f)_L 
      \end{matrix}$ 
      & 
      $\displaystyle \begin{matrix}
        \U(\Nfhalf)_{u_R}\times \U(\Nfhalf)_{u_L} \qquad 
        \\
        \qquad \times \U(\Nfhalf)_{d_R}\times \U(\Nfhalf)_{d_L}
        \\
        \to \U(\Nfhalf)_{u_R+d_L}\times \U(\Nfhalf)_{u_L+d_R}
      \end{matrix}$
      \\\hline 
      $\mui\ne 0$,~$\mu_q\ne 0$ 
      & 
      $\displaystyle \begin{matrix}
        \U(\Nfhalf)_{u_R}\times \U(\Nfhalf)_{u_L} \quad 
        \\
        \quad \times \U(\Nfhalf)_{d_R}\times \U(\Nfhalf)_{d_L}
        \\
        \to \HH ~~\text{(see table~\ref{tb:2color2})}
      \end{matrix}$
      & 
      same as $\mui\ne 0$,~$\mu_q= 0$
      \\\hline 
    \end{tabular}
  \end{center}
  \vspace{-.5\baselineskip}
  \caption{Comparison of the patterns of spontaneous symmetry breaking
    in two-color QCD and in QCD with $N_c\geq 3$ with quark and isospin
    chemical potential in the chiral limit ($m=0$).  $N_f$ is assumed to
    be even.  In the lower two rows the axial anomaly is ignored, as it
    is irrelevant at high density.  In the bottom row, $\mu_q$ is
    assumed to be much smaller than the other scales (e.g., $\mui$ and
    $\Delta$) so that $\mu_q\ne 0$ does not disrupt the condensate at
    $\mu_q=0$. }
  \label{tb:2color1}
  \vspace{5pt}
  \begin{center}
  \begin{tabular}{|c||c|c|}
    \hline 
    & Residual symmetry ($\HH$) & Sign problem 
    \\\hline\hline 
    $N_f=4,8,\dots$ & $[\Sp(N_f/2)]^4$ 
    & absent 
    \vspace{\baselineskip}
    \\\hline
    $N_f=2,6,\dots$ & $\big[\Sp\big(\frac{N_f-2}{2}\big)\big]^4 \times [\U(1)]^2$ 
    $\rule{0pt}{13pt}_{\rule{0pt}{10pt}}$ \!\!\!\! & present 
    \\\hline 
  \end{tabular}
  \end{center}
  \vspace{-.5\baselineskip}
  \caption{Global symmetries that remain intact after spontaneous
    symmetry breaking in two-color QCD with $\mui\ne 0$ and $\mu_q\ne 0$
    in the chiral limit $(m=0)$. Again, $\mu_q$ is assumed to be much
    smaller than the other scales.}
  \label{tb:2color2}  
\end{table} 

The unique symmetries of two-color QCD can readily be incorporated
into RMT by simply replacing the complex random matrices in
\eqref{eq:Z_rmt} with real random matrices. This prescription was
introduced in chiral RMT at zero density in \cite{Verbaarschot:1994qf}
and later generalized to chiral RMT for two-color QCD at high density
\cite{Kanazawa:2009en,Kanazawa:2011tt,Kanazawa:PhD}.  After applying
this prescription, the mapping of section~\ref{sec:silver} from high
isospin to low baryon density is still valid, and the ensuing analysis
for the stressed-singular-value density and the baryon-number Dirac
spectrum parallels the $N_c\geq 3$ case, although the actual
calculations are technically more difficult \cite{Akemann:2010tv}.

We now briefly highlight some physically distinctive features of
two-color QCD. As we will see shortly, the symmetry-breaking pattern
essentially depends on whether $N_f/2$ is even or odd.%
\footnote{A related discussion may be found in
  \cite[Sec.~VII]{Splittorff:2000mm}.}  As an example for even
$N_f/2$, let us take $N_f=4$ with quarks $\{\bar{u}_1, \bar{u}_2, d_1,
d_2\}$.  For nonzero $\mu_q$, the Cooper pairing between $\bar{u}$ and
$d$ becomes energetically costly, so the dominant pairing channels are
$\langle\bar{u}_{1i}\bar{u}_{2i}\rangle$ and $\langle
d_{1i}d_{2i}\rangle$ with $i=R, L$. (Note that these condensates are
color singlets for $N_c=2$.)  Thus in this case the unbroken global
symmetry that leaves these condensates unchanged is $[\Sp(2)]^4$.  It
generalizes to $[\Sp(N_f/2)]^4$ for general even $N_f/2$, as given in
table~\ref{tb:2color2}.  We note in passing that the
high-isospin-density Silver Blaze phenomenon does not occur in this
case, as the NG modes respond to any small $\mu_q\ne 0$ right away ---
the pionic condensate transmutes into the diquark condensates, in a
way analogous to two-color QCD at low baryon density where the chiral
condensate transmutes into the diquark condensate \cite{Kogut:2000ek}.

Next we move on to odd $N_f/2$, focusing on $N_f=2$ and $N_f=6$ for
illustration.  For $N_f=2$ and at large $\mui$, the condensate
$\langle\bar{u}\gamma_5 d\rangle$ forms and persists until $\mu_q$
reaches a threshold $\mu_q^c\sim \Delta/\sqrt{2}$ (see
\cite{Fukushima:2007bj,Andersen:2010vu} for detailed model analyses of
this transition), while at the same time the quark-number density
remains zero at $T=0$, exhibiting the high-isospin-density Silver
Blaze phenomenon. This theory shows essentially the same behavior as
QCD for $N_c\geq 3$. The unbroken symmetry is $[\U(1)]^2$, one of
which is the quark-number symmetry and the other is a rotation
generated by $\gamma_5 I_3$, with $I_3$ the third isospin generator.

For $N_f=6$, $\mu_q\ne 0$ tries to split the coincident Fermi levels
of $\{\bar{u}_{1,2,3},~d_{1,2,3}\}$ to two levels, one each for
$\{\bar{u}_{1,2,3}\}$ and $\{d_{1,2,3}\}$. However,
$\{\bar{u}_{1,2,3}\}$ or $\{d_{1,2,3}\}$ alone involve an odd number
of flavors and cannot support an isotropic BCS pairing by
themselves. Then it would be energetically more preferable to pair as
$\langle\bar{u}_1 \bar{u}_2\rangle$, $\langle d_1 d_2\rangle$, and
$\langle\bar{u}_3 d_3 \rangle$ (up to trivial permutations).  The last
pairing is stressed by $\mu_q$. The residual symmetry in this phase is
the product of $[\Sp(2)]^4$, which leaves $\langle\bar{u}_{1i}
\bar{u}_{2i} \rangle$ and $\langle d_{1i} d_{2i} \rangle$ ($i=R, L$)
unchanged, and $[\U(1)]^2$, which acts on $\bar{u}_3$ and $d_3$ in the
same way as in the $N_f=2$ case. The symmetry for general odd $N_f/2$
is given in table~\ref{tb:2color2}.

The emergence of the sign problem at $\mu_q\ne 0$ also depends on
whether $N_f/2$ is even or odd.  Since the fermion determinant in
two-color QCD is real, the path-integral measure in \eqref{eq:z2color}
for even $N_f/2$ is nonnegative definite, and therefore no sign
problem arises.%
\footnote{The sign problem returns if $\dMu$ or the quark masses are
  made flavor asymmetric \cite{Akemann:2010tv}.}  The
stressed-singular-value density is a smooth function over the complex
plane, unlike for $N_c\geq 3$ where $\mu_q\ne 0$ inevitably causes
strong oscillations (recall figure~\ref{fg:stresseddensity}).  In
contrast, for odd $N_f/2$, the sign fluctuation of the determinant in
\eqref{eq:z2color} is not completely canceled at $\mu_q\ne 0$.
Combining the mapping from high isospin to low baryon density in
section~\ref{sec:silver} with the exact spectral densities in
two-color QCD at low baryon density \cite{Akemann:2010tv} we learn
that the stressed-singular-value spectrum at high isospin density
should exhibit a domain of strong oscillations just as depicted in
figure~\ref{fg:stresseddensity}.  A quantitative study of this
phenomenon in two-color QCD is an interesting future direction.

\section{Concluding remarks}
\label{sec:conclusion}

In this paper we have studied QCD with large isospin chemical
potential $\mui$ for an arbitrary even number of flavors, allowing for
a small mismatch of chemical potentials for different flavors.  In
section~\ref{sec:LEET} we have systematically constructed the
low-energy effective theory of Nambu-Goldstone modes which emerge from
the symmetry breaking due to the BCS pairing of $\bar{u}$ and $d$
quarks.  After formulating the $p$-expansion for coincident Fermi
surfaces, we have extended the scheme to the case where the BCS
pairing is stressed by small $\mu_q\ne 0$, by utilizing the invariance
of the high-isospin-density effective theory under a spurious temporal
gauge transformation involving $\mu_q$.  We also established counting
rules for the $\epsilon$-expansion at high isospin density and
constructed the low-energy effective theory in the leading order of
this expansion.  Using this effective theory we have estimated the
severity of the sign problem showing that, with nonzero stress, the
average sign factor becomes exponentially small for large space-time
volume.  In section~\ref{sec:rmt} we provided a new random matrix
theory that reproduces the finite-volume partition function in the
$\epsilon$-regime.  We introduced ``stressed singular values'' of the
Dirac operator for nonzero stress and showed that the pionic
condensate at large $\mui$ is linked to the near-zero spectrum of the
stressed singular values. Moreover, we found that the microscopic
correlation functions of the stressed singular values in the chiral
limit at large $\mui$ are exactly described by those of the Dirac
eigenvalues at $\mui=0$ and small $\mu_q$, which is a consequence of
an interesting equivalence between our RMT at large $\mui$ and the
conventional one at $\mui=0$ and small $\mu_q$.  This equivalence also
enabled us to elucidate the microscopic mechanism of the
high-isospin-density Silver Blaze phenomenon: the partition function
at $T=0$ is independent of $\mu_q$ although the quark determinant
depends on $\mu_q$. We found that this is due to a rapidly oscillating
part of the stressed-singular-value spectrum.  Intriguingly, this
feature is mathematically the same as for the Silver Blaze phenomenon
at $\mui=0$ and $\mu_q\ne 0$. Furthermore, we pointed out that the
baryon-number Dirac spectrum, i.e., the spectrum of the operator
$\gamma_4(D(\mui)+m)$, can be computed analytically from our new RMT
at least in the chiral limit. The extension of the present work to
two-color QCD was also discussed.

There are many possible future directions. First, the Dirac
eigenvalues with nonzero stress have not been considered in this work.
It is an important but challenging task to compute their microscopic
correlation functions explicitly in the framework of our new
RMT. Second, it would be intriguing to analytically compute the group
integrals of the $\epsilon$-regime partition function
\eqref{eq:Z_qcd_ep} for the general case of nonzero mass and nonzero
stress.  Third, we pointed out that it is possible to obtain the
stressed-singular-value spectrum and the baryon-number Dirac spectrum
in the chiral limit by way of the mapping to low baryon
density. However, this mapping does not work for nonzero quark masses,
and it deserves further study to understand those spectra in the
massive case. Fourth, the meson mass spectrum for nonzero stress can
be determined from the effective theory constructed in this
paper. Fifth, it would be interesting to look into two-color QCD more
thoroughly on the basis of our brief account in
section~\ref{sec:2color}.  Sixth, the extension of this work to the
regime with strong stress ($\mu_q\sim \Delta$) is quite important, but
for us to formulate a low-energy expansion we must first pin down the
correct pattern of symmetry breaking as well as the condensates that
form.  This is not yet fully resolved in dense QCD, and a lot of
elaborate work would be necessary before one can discuss anything
about the spectral properties of the Dirac operator.  
Seventh, the generalization of the RMT in this paper to 
QCD at large baryon chemical potential is an important open problem. 
A salient feature of QCD at high baryon density is that the
Cooper pairing of quarks leads to gauge symmetry breaking, which does
not occur in QCD at high isospin density. 
Despite the fact that the low-energy effective theory of Nambu-Goldstone 
modes at high baryon density is already well known \cite{Alford:2007xm}, 
it is unclear to us how to incorporate a colored condensate into RMT, 
and this obstacle makes it difficult to extend the Dirac 
eigenvalue analysis of the present paper to QCD at high baryon density.  
Last but not least, the results of this paper should be checked in future lattice
simulations.%
\footnote{We mention that numerical simulations of QCD with isospin
  density have already been performed in
  \cite{Kogut:2002tm,Gupta:2002kp,Kogut:2002zg,Kogut:2004zg,deForcrand:2007uz,Cea:2012ev,Detmold:2012wc,Yamamoto:2014lia},
  although the BCS regime of high isospin density seems unexplored
  yet.}  Our analytical predictions are not only of physical
relevance, but also offer a nontrivial benchmark test for any
computational technique that aims to overcome the sign problem.

\acknowledgments

We thank Naoki Yamamoto for his collaboration at an early stage of
this work.  TK was supported by the RIKEN iTHES Project and JSPS
KAKENHI Grants Number 25887014.  TW is supported by DFG (SFB/TRR-55).

\appendix

\section{A potential ambiguity in the effective theory}
\label{app:red}

Let us investigate what happens if we do not the impose the condition
\eqref{eq:GF}.  In the following we denote the effective Lagrangian in
\eqref{eq:L_eff_full} by $\mL_\text{eff}(\mui;\dMu_u,\dMu_d)$.  As a
simple example, consider the choice
\begin{equation}
  \label{eq:shift}
  \dMu_u = -\delta\dMu \quad\text{and}\quad
  \Mu_d = \delta\dMu \quad\text{with}\quad 
  \delta\dMu=\delta\dmu\1_{\Nfhalf}\,,
\end{equation}
where $\delta\dmu\ll\mui$.  This is equivalent to shifting
$\mui\to\mui+\delta\dmu$ as noted in section~\ref{sec:prelim}.  We now
encounter an ambiguity since the effective Lagrangian could be written
as $\mL_1=\mL_\text{eff}(\mui+\delta\dmu;0,0)$ or
$\mL_2=\mL_\text{eff}(\mui;-\delta\dMu,\delta\dMu)$.  In the former
case all low-energy constants are evaluated at $\mui+\delta\mu$ and
the correction terms in \eqref{eq:UV} are absent, while in the latter
case all low-energy constants are evaluated at $\mui$ and the
correction terms in \eqref{eq:UV} are present.  Although the
underlying microscopic theory is the same, it is not obvious that
$\mL_1$ and $\mL_2$ are identical.  Indeed, they need not be identical
but can differ by terms that are of higher order in the $p$-expansion.

To understand this ambiguity in a simpler setting, let us turn to the
effective theory for relativistic $\U(1)$ superfluids
\cite{Son:2002x}. At high density, where the interaction is weak, the
leading-order effective Lagrangian in Minkowski space for the $\U(1)$
NG mode $\phi$ is given by
\begin{align}
  \label{eq:L_U1}
  \mathcal L^{(2)}_\text{eff}(\phi) = 
  \frac{N_cN_f}{2\pi^2}\mu^2 \Big[(\del_0 \phi)^2 - \frac{1}{3}
  (\del_i \phi)^2  \Big] . 
\end{align}
If $\mu$ is increased to $\mu+\mu'$ with $\mu'\ll \mu$, the factor
$\mu^2$ in \eqref{eq:L_U1} is merely replaced by $(\mu+\mu')^2$ to
give a new $\mathcal L^{(2)}_\text{eff}(\phi)$. On the other hand,
according to the same kind of spurion analysis as in
section~\ref{sec:LEET}, $\mathcal L^{(2)}_\text{eff}(\phi)$ should be
invariant under a time-dependent $\U(1)$ symmetry, under which
$\phi\to\phi+\alpha$ and $\mu'\to \mu'+\del_0\alpha$. Thus the effect
of $\mu'$ can be incorporated via the prescription $\del_0\phi \to
\del_0\phi - \mu'$, which yields
\begin{align}
  \mathcal L^{(2)}_\text{eff}(\phi) = \frac{N_cN_f}{2\pi^2}\mu^2  \Big[  (\del_0 \phi - \mu')^2 - \frac{1}{3}(\del_i \phi)^2  \Big] 
  \ne 
  \frac{N_cN_f}{2\pi^2} (\mu+\mu')^2  \Big[  (\del_0 \phi)^2 - \frac{1}{3}(\del_i \phi)^2  \Big] .
\end{align}  
This discrepancy stems from the fact that higher-order terms in the
full Lagrangian $\mathcal L_\text{eff}$ were discarded. If we look at
the full effective theory derived by Son \cite{Son:2002x},
\begin{align}
  \mathcal{L}_\text{eff}(\phi) = \frac{N_cN_f}{12\pi^2} \big[ (\del_0\phi-\mu)^2-(\del_i\phi)^2 \big]^2 \,,
\end{align}
we can easily see that the two prescriptions $\mu\to\mu+\mu'$ and
$\del_0\phi\to \del_0\phi-\mu'$ do give an identical expression.
However, if the effective theory is truncated at some order, one in
general ends up with two expressions that differ by higher-order
terms. We stress that this poses no problem at all as long as $\mu'$
is so small that its higher-order terms can be safely
neglected. However, if $\mu'$ is not a small parameter, the
prescription $\del_0\phi\to \del_0\phi-\mu'$ can no longer be applied
because the Taylor expansion of $\mathcal L_\text{eff}$ in $\mu'$
would not be convergent. Then one is forced to start from an effective
theory defined at $\mu +\mu'$.
   
The lesson from this simpler example also applies to our effective
theory $\mL_\text{eff}(\mui;\dMu_u,\dMu_d)$.  In our case the
ambiguity can be avoided if we make sure that $\mui$ is not shifted,
which is guaranteed if we impose the condition \eqref{eq:GF}.

\bibliography{draft_v3.bbl}
\bibliographystyle{JBJHEP}
\end{document}